\documentclass[useAMS,usenatbib]{mn2e}

\usepackage{graphicx}
\usepackage{amsmath}
\usepackage{amssymb}
\usepackage{natbib}

\newcommand{\apj}{Ap.J.}
\newcommand{\aj}{A.J.}
\newcommand{\mnras}{MNRAS}
\newcommand{\aap}{A\&A}
\newcommand{\pasp}{PASP}
\newcommand{\fr}{Fregeau et al.}
\newcommand{\hth}{Paper\ I}

\newcommand{\add}{ }

\newcommand{\be}{\begin{equation}}
\newcommand{\ee}{\end{equation}}

\def\trh0{t_{rh}(0)}

\def\apgt{\ {\raise-.5ex\hbox{$\buildrel>\over\sim$}}\ }
\def\aplt{\ {\raise-.5ex\hbox{$\buildrel<\over\sim$}}\ }

\title[Star Cluster Evolution with Primordial Binaries II]
{Star Clusters with Primordial Binaries:\break
II. Dynamical Evolution of  Models in a Tidal Field}
\author[M. Trenti, P. Hut and D.C. Heggie]{Michele Trenti$^{1,2}$\thanks{E-mail
addresses:
trenti@stsci.edu (MT); piet@ias.edu (PH); d.c.heggie@ed.ac.uk (DCH)}, 
Douglas C. Heggie$^{3}$\footnotemark[1] and Piet Hut$^{4}$\footnotemark[1]\\
$^{1}$Space Telescope Science Institute, 
3700 San Martin Drive,
Baltimore, MD 21218, USA\\
$^{2}$Yukawa Institute for Theoretical Physics, Kyoto University,
606-8502 Kyoto, Japan\\
$^{3}$School of Mathematics,
	University of Edinburgh,
	King's Buildings, Edinburgh EH9 3JZ,
	Scotland, U.K. \\
$^{4}$Institute for Advanced Study, Princeton, NJ 08540, USA}

\begin{document}

\date{Accepted ; Received ; in original form }

\pagerange{\pageref{firstpage}--\pageref{lastpage}} \pubyear{2006}

\maketitle

\label{firstpage}

\begin{abstract}
We extend our analysis of the dynamical evolution of simple star
cluster models, in order to provide comparison standards that will aid
in interpreting the results of more complex realistic simulations.  We
augment our previous primordial-binary simulations by introducing a
tidal field, and starting with King models of different central
concentrations.  We present the results of $N$-body calculations of
the evolution of equal-mass models, starting with primordial binary
fractions of 0 - 100 \%, and $N$ values from 512 to 16384.  We also
attempt to extrapolate some of our results to the larger number of
particles that are necessary to model globular clusters.

We characterize the steady-state `deuterium main sequence' phase in
which primordial binaries are depleted in the core in the process of
`gravitationally burning'.  In this phase we find that the ratio of
the core to half-mass radius, $r_c/r_h$, is similar to that measured
for isolated systems (Heggie et al. 2005).  In addition to the
generation of energy due to hardening and depletion of the primordial
binary population, the overall evolution of the star clusters is
driven by a competing process: the tidal dissolution of the system. If
the primordial binary fraction is greater than $5 \%$ and the total
number of particles $N \geq 8192$, we find that primordial binaries
are not fully depleted before tidal dissolution, in systems initially
described by a King model with a self-consistent tidal field.

We compare our findings, obtained by means of direct $N$-body
simulations but scaled, where possible, to larger $N$, with similar
studies carried out by means of Monte Carlo methods (Fregeau et
al. 2003, 2005).  We find significant qualitative and quantitative
differences with the results in the earlier paper.  
Some of these differences are explicable by the
different treatment of the tidal field in the two studies. Others,
however, confirm the conclusion of Fregeau et al (2005) that the
efficiency of binary burning in the earlier Monte
Carlo runs was too high. There remain unexplained differences,
however.  In particular, the  binary population appears to be
depleted too quickly, even in the most recent Monte Carlo results.

\end{abstract}

\begin{keywords}
globular clusters -- methods: $N$-body simulations -- stellar dynamics.
\end{keywords}

\section{Introduction}

Several observational surveys of globular clusters have highlighted
the presence of binaries, which are found to constitute up to $50
\%$ of the mass of the core \citep[e.g.,
see][]{rub97,alb01,bel02,pul03}.  Such a high number of objects is
extremely unlikely to have been formed dynamically, as, at the
density required to create a significant binary population due to
tidal capture and three body encounters, the most probable outcome
from these processes is a merger \citep{che96,pzm02}, which could
eventually lead to the formation of a central intermediate mass black
hole \citep{pz1}.  Therefore these binaries are regarded as being {\sl
primordial}.
% {\bf expand with something about exotic products in GCs
%as blue straggers, ... ?}
The {primordial} binary fraction appears to be a key parameter for
determining the dynamical evolution of a star cluster, as binaries are an
efficient heating source which can halt core collapse. In addition,
interesting astronomical objects can be produced due to three- and
four-body encounters, such as blue stragglers, X-ray binaries and
binaries containing pulsars.

The evolution of star clusters with primordial binaries still presents
open issues, mainly due to the intrinsic complexity of numerical
simulations of a system where the local dynamical timescale may be
many orders of magnitude smaller than the global relaxation timescale.
(Hard binaries have an orbital period of a few hours, while the half
mass relaxation time $t_{rh}$ may be of order a few billion years.)  Numerical
simulations have thus been performed either using approximate
algorithms such as Fokker Planck or Monte Carlo methods
\citep{gao91,gie00,fre03}, which often have to rely on a knowledge of
the relevant interaction cross sections; or using direct $N$-body
codes but employing only a modest number of particles of order $N \approx 10^3$
\citep{mcm90,mcm94,heg92}. Both approaches could potentially lead to
misleading results. On the one hand Monte Carlo methods rely on input
physics that may not accurately reflect the realistic interactions in
a densely populated and fluctuating core (e.g. the velocity
distribution may be assumed to be isotropic), while interaction cross
sections in the case of unequal masses are not well known.
On the other hand, extrapolation from direct
simulations may prove to be non trivial, as it is not clear how to
scale the results obtained with the number of particles.

We have thus recently started (\citealt{heg05}, hereafter \hth) a
program to survey the basic properties of the evolution of an
idealised star cluster with a population of primordial binaries, with
the aim of placing  a stepping stone between simplified models, such as
those of Fokker-Planck type, and 
realistic complex numerical simulations, such as those recently
performed by \citet{por04b}. In the first paper of our series we have
focused on the evolution of isolated, equal-mass models, with a
primordial binary fraction in the range $0-100~\%$ and with a number
of particles $N$ in the range from $256$ to $16384$. We have compared
our results not only with those of the study carried out by
\citet{gao91} (who used a Fokker Planck code), and also to some extent
with \citet{gie00} and with \citet{fre03}, but also with a
theoretical model which predicts the evolution of the core radius
\citep{ves94}. We have shown that significant differences arise between
direct $N$-body simulations and the Fokker Planck calculations of
\citet{gao91}. In particular we found that, starting from the same
initial conditions, the number of binaries in the core after the core
collapse is significantly lower in our simulations than the number
reported by \citet {gao91}. This could have important consequences if
one attempted to infer the original primordial binary fraction by
observations of the current number of binaries in the cores of
globular clusters. Another interesting finding is that the ratio of
core radius $r_c$ to half mass radius $r_h$ in the post collapse phase seems not to follow
the theoretical expectation from the work of \citet{ves94}, as, by
varying the number of particles $N$, we observed a steeper decrease of
$r_c/r_h$ than that expected ($\propto 1/\log{(0.4 N)}$).

In this paper we extend our simulations to include the effect of the
tidal field of the galaxy.  We consider the evolution of King models
with different values of $N$, the concentration index $W_0$ and the
primordial binary ratio $f$. We
compare our direct simulations to the Monte Carlo simulations performed by
Fregeau et al (2003, 2005), hereafter \fr~ They studied the evolution of isolated
and tidally truncated star clusters with a population of primordial
binaries in the range $2-20~\%$ and a realistic particle number
($N=3~10^5$). Their two-dimensional Monte Carlo Method is expected to
offer significant advantages over the one-dimensional code used by
\citet{gao91} as the influence of anisotropy in the velocities can be
taken into account, and it is unnecessary to assume that the
distributions of energy and position (for binaries) are independent.  The work of \fr~ led to some important
conclusions.  For tidally limited clusters they noted for the first
time the possibility of an initial expansion of the core radius of the
cluster in the presence of a significant population of primordial
binaries, when starting from models with a high central concentration
(i.e.  King profiles with $W_0 \gtrsim 7$). \fr~also showed that, in general,
primordial binaries delay the deep core collapse phase that is
observed in clusters with only single stars, so that the system can be
tidally dissolved before collapsing. The comparison of our results
with their simulations is mainly focused on fundamental quantities
like the disruption rate of the binaries, the core radius and the
dissolution time-scale.  {It must be noted, however, that Fregeau et
al improved their code between the 2003 and 2005 papers by the inclusion of direct
numerical integration of encounters.  They note \citep{fre05} that
this alters some of their earlier results, as we mention at
appropriate points in the present paper.  In particular \citet{fre05} note that binary burning was too efficient
in their earlier models.}

The paper is organized as follows: in the next Section we present the
parameters of our simulation dataset and the numerical method used. In
Sec. 3 we give a physical picture of the evolution of a star cluster
with primordial binaries and tidal field that will guide our
interpretation of the simulations. In Sec.~\ref{sec:fregeau} and 5 {we
present our results for runs with a tidal field and a tidal cutoff, respectively, including} comparison with
the results of \fr~ The last section of the paper provides a summary and
discussions.

\section[]{Simulations: setup and analysis}\label{sec:sim}

\subsection[]{Specification of the models}

%{\bf I divided this section into subsections, because new material was
%  brought forward from the next section (definition of core radius). -
%  D.}

The models considered in this paper are tidally limited, with stars of
equal mass $m$. The initial distribution is a King model with
scaled central potential  $W_0=3,7,11$. Our standard models have a
primordial binary ratio of $10-20~\%$, although we have also performed some
runs with fewer or no primordial binaries { ($0,2,5$\%)} as well as with higher ratios,
$50~\%$ and $100~\%$. As in \hth~ we define the primordial binary
fraction as:
%%%%%
\begin{equation}\label{eq:f}
f = n_b/(n_s+n_b)
\end{equation}
%%%%
with $n_s$ and $n_b$ being the number of singles and binaries
respectively.  This implies that the fraction of the total mass in
binary stars, in the case of equal masses, is larger in the following
way:
$$
f_m = 2n_b/(n_s+2n_b)
$$
For example, for a run with 10\% primordial binaries, $f=10\%$ whereas
$f_m=2/11\sim 18\%$.

All our results are presented using standard units \citep{hm86}
in which
{\add $$
G = M = - 4 E_{\rm T} = 1
$$ } where $G$ is the gravitational constant, $M$ is the total mass, and
$E_{\rm T}$ is the total energy of the system of bound objects.  In
other words, $E_{\rm T}$ does not include the internal binding energy
of the binaries, only the kinetic energy of their centre-of-mass
motion and the potential energy contribution where each binary is
considered to be a point mass.  We denote the corresponding unit of
time by 
\begin{equation}\label{eq:td}
t_d = GM^{5/2}/(-4E_T)^{3/2}
\end{equation}
{in general}. For the
relaxation time, we use the following expression
\begin{equation}\label{eq:trh}
t_{rh} = \frac{0.138 N r_h^{3/2}}{\ln{(0.11 N)}},
\end{equation}
where $r_h$ is the half-mass radius, and $N$ denotes the number of original objects, i.e.
$N=n_s+n_b$.  When we discuss a run with $N=4096$ and 50\% primordial
binaries, we are dealing with $6144$ stars.
We have considered runs with $N$ in the range
$512-16384$. 

Our runs follow the evolution of the system until  the number of particles drops below $10$. However,
to avoid possible biases introduced by long-living small-$N$
configurations, we define the dissolution time $t_{dis}$, as the moment
when $98~\%$ of the initial mass of the system is no longer bound.

Following \fr, our initial distribution of the binaries' binding
energies is flat in log scale in the range from $\epsilon_{min}$ to
$133\epsilon_{min}$, with $\epsilon_{min}=m \sigma_c(0)^2$.  Here
$\sigma_c(0)$ is the initial central velocity dispersion and this
choice, if applied to an isolated Plummer model,  corresponds approximately
to the standard range adopted in \hth, i.e. to $\epsilon_{min} \approx
5.1~kT$ and $133\epsilon_{min} \approx 680~kT$, where
the mean kinetic energy per particle (over the entire system, with
binaries replaced by their centres of mass) is $3kT/2$.

To follow the evolution of the system we have used Aarseth's NBODY6
\citep{aar03}, which has been slightly modified to provide additional
runtime diagnostics on the spatial distributions of single and
binary stars.
%  {\bf [Remainder of this paragraph moved here from old
%Sec. 4.1 (now defunct). - D]}   
A King model with $W_0=7$ required almost one
month of CPU time on a Pentium4 3Ghz PC before tidal dissolution for
$N=16384$ and $f=10\%$; the use of GRAPE hardware would not help
significantly, as the computational bottleneck is in the treatment of
the local dynamics of binaries \citep[however see][for a discussion on
the scaling of the computational time with the number of particles in
the presence of primordial binaries]{mak90}. 

The galactic tidal field is treated in our standard sample of
simulations as that due to a point mass, and the tidal force acting on each
particle is computed using a linear approximation of the
field.
The tidal radius $r_t$ is defined as \citep{spi87}:
%%%%%%%%%%
\be\label{eq:rt}
r_t^3 = \frac{M}{3M_{Gal}}R_{Gal}^3,
\ee
%%%%%%%%%%
where $M_{Gal}$ is the galaxy mass, and $R_{Gal}$ is the distance of
the centre of the star cluster from the centre of the galaxy.  { We
assume $R_{Gal}$ is constant, i.e. the cluster describes a circular
orbit.}  Thus the equations of motion of a star in the cluster are
\begin{equation}\label{eq:tidalFORCE}
\left.\begin{array}{rl}
  \ddot x + 2 \omega \dot y + (\kappa^2 - 4 \omega^2) x  & = 
  - \phi_x  \\
  \ddot y - 2 \omega \dot x \phantom{\ + (\kappa^2 - 4\omega^2)x}
  & =  - \phi_y \\
  \ddot z \phantom{+ 2 \omega\dot y + \kappa^2 - 4} + \nu^2 z &
  =  - \phi_z,
\end{array}\right\}
\end{equation}
where the axes are orientated conventionally with respect to the
direction to the centre of the galaxy and the galactic motion of the
star cluster, $\omega$ is its angular velocity, $\kappa$ is its
epicyclic frequency, $\nu$ is its vertical frequency, and the right
sides are the components of the gravitational acceleration due to the
other stars.  The value of $r_t(0)$ is set at the beginning of the
simulation to match the cut-off given by the adopted initial King
model.  In our standard runs particles are removed from the system
when they reach a distance from the density centre greater than twice
the instantaneous tidal radius $r_t$, and the value of $r_t$ is
updated during the simulation according to the decrease of $M$, the
total mass of the system.
%{\bf
%  [Michele, please check that this is what your version of the code
%    did.  Is the total mass the mass inside $r_t$ or inside $2r_t$?]}
%%% This is correct. The mass is initially set within r_t, but a
%%% particle is removed only at 2r_t

Note that this method of treating the influence of the galactic field
is different from the one adopted by \fr, as they imposed a radial
``cutoff'', i.e. they did not consider the tidal force acting on
individual particles, and simply removed particles whose distance
exceeded $r_t$ \citep[see the description of their code
in][]{jos00}. {\add At variance with our tidal field treatment, the
simulations with tidal cut-off do not include the Coriolis and
centrifugal contributions to the particles acceleration. Therefore the
two approaches have important physical differences.} The effects of this
different treatment are discussed in Sec.~\ref{sec:tidalcut}, where we
describe a subset of simulations ($W_0=7$ and $W_0=3$, $f=10\%$, with
$N$ from $512$ to $16384$) using the tidal radial cutoff treatment
adopted by \fr~

Our results, unless otherwise noted, are presented by applying a
triangular smoothing filter with width $2.5~t_{rh}(0)$ (see Fig.~3
and Sec.~3.2 in \hth~ for further details).

A summary of the runs is presented in Table~\ref{tab:runs}.

%%%%%%%%%%%%%%%%%%%%%%%%%%%%%%

\subsection{Core radius: definition}\label{sec:rcDEF}

%{\bf This subsection moved here from old Sec.4.3} 

\begin{figure}
\resizebox{\hsize}{!}{\includegraphics{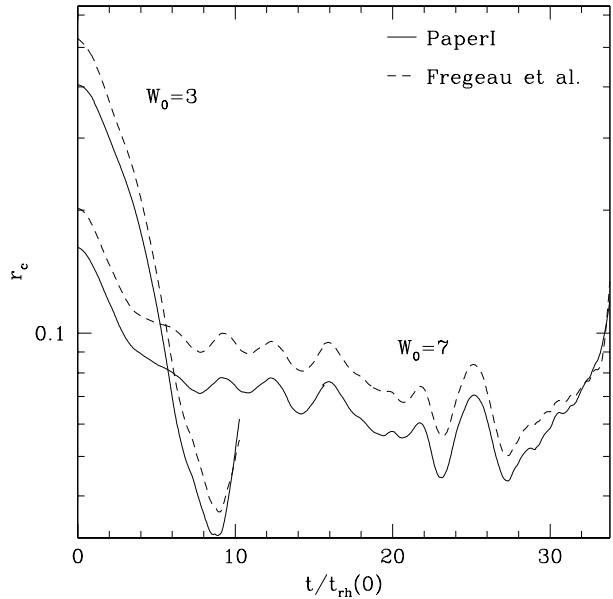}} \caption{Core radius
 measured using the method in \hth~(Eq.~\ref{eq:rc}) and in
 \fr~(Eq.~\ref{eq:rcFR}) for runs with $N=16384$ and $f=10\%$. Note that
 the density-averaged definition of the core radius (\hth)
 underestimates the core radius at the beginning of the simulation
 compared with
 the theoretical value for the corresponding King model (i.e. $0.67$
 and $0.21$ for $W_0 = 3$ and $7$, respectively). In
 this paper we adopt the definition used by \fr~ for the core
 radius.}\label{fig:core}
\end{figure}

In \hth~we adopted the following definition for the core radius:
%%%%
\be \label{eq:rc}
\tilde{r}_c= \sqrt{\frac{\sum_{i=1,\tilde{N}}{r_i^2 \rho_i^2}}{\sum_{i=1,\tilde{N}}{\rho_i^2}}},
\ee
%%%%
where the sum is made over the particles within the half mass radius
and the density $\rho_i$ around each particle is computed from the
distance to the fifth nearest neighbour \citep{cas85}.
This definition presents some systematic differences (see
Fig.~\ref{fig:core}) from the one adopted by \fr:
%%%%
\be \label{eq:rcFR}
r_c=\sqrt{\frac{3 \sigma_c^2}{4 \pi G \rho_0}},
\ee
%%%%
where $\sigma_c^2$ is the central velocity dispersion {\add (mass weighted
as in \citealt{ves94})} and $\rho_0$ the central density \citep{spi87}.
For a proper comparison between our set of simulations and the runs
discussed by \fr~ it is important to refer to the same quantities, and
in this paper we adopt as definition for the core radius $r_c$ the
choice taken by \fr, i.e. Eq.~(\ref{eq:rcFR}). We estimated the
velocity dispersion $\sigma_c^2$ using all the particles within the
radius defined by Eq.~(\ref{eq:rc}), while $\rho_0$ has been obtained
from the Lagrangian radius containing $1~\%$ of the instantaneous mass
of the system.  For the purpose of comparison with \hth, we also
report in Table~\ref{tab:runs} the value of the alternative definition
for the core radius, $\tilde{r}_c$, at the end of the core collapse.

%%%%%%%%%%%%%%%%%%%%%%%%%%%%%%
%%%%%%%%%%%%%%%%%%%%%%%%%%%%%%

\subsection{Time-scales}

In the presence of primordial binaries and a tidal field, several
dynamical processes operate, each with its own time scale.  Here we
summarize them:
\begin{itemize}
\item {\emph{ Dynamical time} ($t_{d}$, cf. eq.(\ref{eq:td}))}: within
a small numerical factor, this is the typical timescale
  for a particle to cross the system.  It also represents the timescale on
  which approximate dynamical equilibrium is established in the case
  of initial conditions starting out of equilibrium. (Note however that all our
  simulations start from approximate dynamical equilibrium.)
\item {\emph{ Half-mass relaxation time} ($t_{rh}$, cf. eq.(\ref{eq:trh}))}: this is the
  timescale for energy diffusion in phase space, i.e. for establishing
  thermal equilibrium. It is of order $N/\log(0.11 N)~t_d$.
\item {\add {\emph{Total-mass relaxation time} ($t_{rt}$)}: this is the
  timescale for relaxation of the cluster as a whole and is defined
  using the tidal radius as characteristic length scale for estimating
  the relaxation time. $t_{rt}$ is proportional to $t_{rh}$ (defined
  above): $t_{rt}/t_{rh} \propto (r_t/r_h)^{3/2}$.}
\item {\emph {Time for initial core contraction} ($t_{cc}$)}: this is the
  timescale on which the core of the cluster contracts (or expands) to
  reach a quasi-steady configuration in which the energy released in
  the core fuels the expansion of the half mass radius. The typical value of
  $t_{cc}$ is of the order of $10 \trh0$.
\item {\emph{Tidal dissolution time} ($t_{dis}$)}: Time needed to lose
  $98\%$ of the initial mass of the system due to escape across the tidal
boundary. Depending on the tidal field strength, the initial
  concentration of the system and the initial number of particles, this timescale
  ranges from a few relaxation times to a few tens of relaxation
  times.
\item {\emph{ Tidal half-mass dissolution time} ($t_{Half}$)}: Similar to
  $t_{dis}$, but defined as the time needed for  $50\%$ of the mass of the
  system to escape.
\item {\emph{ Binary burning time} ($t_{bb}$)} Time needed to deplete
  $80\%$ of the initial number of binaries. Around this time
  gravothermal oscillations set in, provided that $N$ is large
  enough. For initial binary fractions above $10\%$, $t_{bb}$  is of the
  order of several tens to hundreds of initial relaxation times.

\end{itemize}

\section{The physical picture: primordial binaries and the tidal field}\label{sec:picture}

The evolution of an  isolated star cluster made of equal mass
stars is driven initially by the gravothermal instability. Two-body encounters
drive a heat flow from the central region of the star cluster, which
behaves like a self gravitating system with negative specific heat, to
the halo. This triggers a thermal ``collapse'' on the timescale of the
heat flow process.

The ``collapse'' phase lasts until an efficient form of energy
generation in the centre can stop the process by providing an energy
production rate equal to the energy loss rate by two body relaxation
from the region of the core. In the case of systems consisting of
single stars only, the collapse is stopped only after one or more
binaries are formed due to three body encounters in the dense central
region.  The collapse typically takes a time $t_{cc}$ of order $15~t_{rh}(0)$, where
$t_{rh}(0)$ is the initial half-mass relaxation time.  At this point,
i.e. at the end of core collapse (sometimes called ``core bounce''),
the central density can exceed $10^4$ times the average density
inside the half-mass radius $r_h$ (for $N=16384$, or far more for
larger N).
%% {\bf Small paragraph currently commented out (v2), but probably to be
%% reintroduced later}
%---------------------------------------------------------------------
% PIET SUGGESTS TO QUANTIFY BETTER THE NEXT PARAGRAPH (v1);
% commented out in v2; to be discussed and reintroduced in later versions
%---------------------------------------------------------------------
% Piet's comment is that 
%  (1) there may be more than one binary in the core
%  (2) gravothermal oscillations are tricky, e.g. we need N large enough
%
%
%-------------- 
%The typical evolution then proceeds with sudden oscillations in the
%size of the core (the gravothermal oscillations, \citealt{mak96}), as
%it may happen that one of these dynamically formed binaries is kicked
%out from the central region, so that the system collapses again until
%a new binary is formed.
After core bounce, the generation of energy in the core is accompanied
by a steady expansion of the half mass radius.  If $N$ is large
enough, however, binary activity in the core may cause a temporary
temperature inversion which drives the gravothermal instability in
reverse.  This suppresses binary activity, and further collapses may
recur, in a succession of ``gravothermal oscillations''
(\citealt{1983MNRAS.204P..19S}).

%%  of the
%% system steadily expands, at a rate that is set by the energy
%% production in the core.

This picture is strongly modified when a population of primordial
binaries is present, as the generation of energy due to existing
binaries is more efficient than that due to dynamically formed pairs
(see Sec.~2 of \hth). For this reason, if the primordial binary
fraction is above a few percent, the initial contraction of the core
is more gentle, and it is halted at a lower core density (and a larger
core radius).{\add The timescale for initial core contraction ($t_{cc}$)
remains of the same order of magnitude as} in the case where no
binaries are initially present (see Fig.~{ 17} in \hth).  From then
until the time when the population of binaries becomes heavily
depleted, the evolution of the system proceeds with an almost constant
ratio of core radius to half mass radius.  This phase lasts for a time
proportional to the initial binary ratio, and thereafter a phase of
gravothermal oscillations sets in, much as for the case where there
are no primordial binaries (see Figs.~{ 5-6} in \hth).  For an
isolated system, provided that $N$ is large enough (i.e. $N \gtrsim
10^4$), the onset of gravothermal oscillations occurs when the mass of
binaries has fallen to roughly 10--15\% of its initial value
({\citealt{fre03}}, Fig.4; \hth, Figs. { 5,7}).  These are empirical
results, however, and may change for primordial binary fractions
smaller than the lowest used in these simulations, i.e. 2\%.

The presence of a tidal field introduces a new timescale for the
evolution of the system, i.e. $t_{dis}$. In the presence of a tidal field, the star
cluster steadily loses mass at a rate that is almost independent of
the properties of the central parts of the system, where binaries
accumulate (because of mass segregation) and act as an energy source
for the system.  In principle the tidal dissolution can be so fast ($t_{dis} < t_{bb}$) 
that the system does not have enough time to deplete its reservoir of
primordial binaries before dissolving.  On the other hand, if the
initial primordial binary fraction is low or if the tidal field is
weak, then the system can burn almost all of the binaries before being
dissolved, and it then undergoes a phase of gravothermal oscillations,
as already described for an isolated system, provided that $N$ is
large enough.

In an isolated system with an initial binary ratio of $10\%$,
approximately $100~t_{rh}(0)$ is the typical timescale $t_{bb}$ required for
disrupting $80\%$ of the binary population considered in our runs
(e.g., see Fig.~{7} in \hth).
Note, however, that the range of binding energy in these runs is
$5-700~kT$, where $3/2 NkT$ is the total kinetic energy of
the cluster (when the binaries are replaced by their barycentres).  The
time to disrupt most of the binaries is certainly dependent on the
initial distribution.

The time to disrupt most of the binaries can also be estimated with the
following argument.  A primordial binary in the range of $5-700~kT$
gives off, on average, around $100~kT$ before leaving the system (or
being dissolved; see e.g., \citealt{hut92}). Thus if we start with a
$10\%$ primordial binary population each single star receives around
$10~kT$. The energy loss at the half mass radius due to the
temperature gradient is of the order of $0.1~kT$ per star per
relaxation time, so we have to wait an order of $100$ relaxation times
before depleting the binaries.  The same argument suggests that the
time taken to burn most of the binaries is proportional to $f$.

% {\bf mt: refer to discussion of the \fr~results on
%evaporation? }

%{\bf I can't see this in
%  Sec.4.1.  We just mention the same statement.  I think it would be
%  great if we could briefly describe a run with f = 5\%, including a
%  figure like 2-5.  There is a good place for it in Sec.4.2, where we
%  talk about these models.}

%\section{Comparison with Fregeau et al.}\label{sec:fregeau}
\section{Tidal field runs: Results and Comparisons}\label{sec:fregeau}

%{\bf Note name change; I think the main purpose of this section should be to present
%  the results, i.e. to state what really happens, and not to focus on
%  the comparison with Fregeau et al, which is liable to be outdated
%  pretty soon.  I also deleted most
%  of the old Sec. 4.1, which repeated a lot of what had already been
%  said.  The rest I moved into Sec.3. - D}

In this Section we discuss the properties of our simulations, our main
purpose being to set out the essential empirical facts about the evolution of systems with primordial
binaries in a galactic tidal field.   We shall also
compare our results with those obtained by \fr, though they treated
the tidal field as a cutoff.  To elucidate this comparison, we were
motivated to carry out further $N$-body runs with the same treatment (i.e. a
cutoff).  We defer presentation of those calculations to the next
section, however;  in the present section we consider only runs that include
our most realistic treatment of the tidal force (eq.(\ref{eq:tidalFORCE})).

\subsection{Total mass and dissolution timescale}\label{sec:tidal}

Our results are summarized by the values of $t_{dis}$ in
Tab.~\ref{tab:runs} and Tab.\ref{tab:tbEv}, and details for a subset of the runs are
presented in the top panel in Figs.~\ref{fig:fr13}-\ref{fig:fr12};
these are intended also to facilitate comparison with corresponding
plots in \fr, as specified in the captions.  The evolution of the
total mass in our different runs depends on three initial parameters:
(i) the initial fraction of primordial binaries, $f$; (ii) the scaled
central potential of the initial King model, $W_0$; and (iii) the
initial number of objects, $N$.  We consider the dependence on each in
turn.

\subsubsection{Dependence on $f$}

The initial fraction of primordial binaries has only a small influence
on the rate of mass loss of the system and on $t_{dis}$.  The
dependence is, however, systematic, as can be seen from
Fig.~\ref{fig:2}.  (Though this figure shows runs with $W_0=7$, the
trend shown is representative of all concentration indexes that we
have studied.)  The simulation with the longest dissolution time is
the one starting with $100\%$ of primordial binaries.  This may not be
surprising, as this run has the highest total number of stars and so, after
disruption of a given proportion of binaries, it has the longest
relaxation time.  On the other
hand it is closely followed by the simulations with low binary content
($2~\%$ and $5~\%$) and then by the run with single stars only, which
is dissolved in $90\%$ of the time needed in the $100~\%$ case.  This
may be due to the fact that simulations with $f \lesssim 5\%$ deplete
their primordial binaries before dissolution ($t_{bb}<t_{dis}$); this leads to a deep
core collapse, which creates a  strongly bound core which is rather resilient to
dissolution.  Curiously, simulations with $20 \%$ of primordial
binaries are subject to the fastest tidal dissolution, and this could
be due to effects related to mass segregation, which is absent if
$f=0$ or $1$, and which tends to expel single stars.

\subsubsection{Dependence on $W_0$}\label{sec:w-dependence}
Consider, as an example, runs with $N = 4096$ and $f = 10\%$.  A low
concentration King model with $W_0=3$ is tidally dissolved in
approximately $13~\trh0$ (Table \ref{tab:runs}), while the system can
survive for approximately $50~\trh0$ for $W_0=7$; but $W_0=11$ is
intermediate between the two ($t_{dis} \approx 25~\trh0$).  This
non-monotonic dependence on concentration index $W_0$ is a general
feature, independent of the number of particles.  It may be understood
in terms of the non-monotonic dependence of the tidal radius on $W_0$,
at fixed total mass and total energy (Tab.\ref{tab:rt}; see also
Fig.~8.3 in \citealt{hh}).  (Another way of looking at this is that,
in the units we adopt, the strength of the tidal field varies
non-monotonically with $W_0$.)  Though the difference in $r_t$ between
the cases $W_0 = 7$ and $W_0 = 11$ may not seem great, two further
factors may be relevant: (i) $r_h$ also increases slightly between
$W_0 = 7$ and $W_0 = 11$ (Table \ref{tab:rt}), which would increase the vulnerability of the
more concentrated system, even if the tidal field were the same; (ii)
high-concentration models begin with a short period of core {\sl expansion}
(Fig.\ref{fig:fr12}), and the associated burst of energy generation
enhances tidal overflow. 

{\add If the tidal dissolution time is expressed in units of the total-mass
relaxation time ($t_{rt}$), then the differences between the different
King models are greatly reduced. By accounting for the factor
$(r_t/r_h)^{3/2}$, so that the time is in units of $t_{rt}$, the
difference in tidal dissolution time between the various King models
considered here is within $20\%$ for $N=4096$ and within $10\%$ for
$N=16384$. 
%This suggests that tidal dissolution of the cluster is
%mainly due to diffusion of particles near the critical escape energy
%(see also \citealt{tf05}).
}

%In the units we adopt, in which the total initial mass and virial radius are
%constant,  the initial tidal radius varies non-monotonically
%with $W_0$ (Tab.\ref{tab:radii};), and therefore so does the strength
%of the tidal field.  

% The tidal field strength is minimal around $W_0
%\approx 8.5$, where $r_t(W_0)$ has a maximum (see, for example,
%; Table {\ref{tab:radii}}). In addition, at the beginning of the run as
%the central concentration index $W_0$ increases, the energy production
%in the centre of the system is more efficient and the expansion of the
%cluster starts earlier, so that tidal dissolution is more
%efficient. The process is strongly non-linear (the tidal radius
%shrinks depending on the remaining mass, see Eq.~\ref{eq:rt})
%explaining the marked difference in the dissolution times between
%$W_0=7$ and $W_0=11$ even if the tidal field strength is not too
%dissimilar in the two cases. Based on this picture we expect that
%$W_0=7$ is close to a maximum in the dissolution time, as for more
%concentrated models the higher central concentration balances the
%effects of a marginally larger tidal radius.

%%%%%%%%%%%%%%%%%%%%%%%%%%%%%%%%%%%%%%%%%%%%%%%%%%%
% table for differences in tidal dissolution time %
\begin{table}\label{tab:radii}
\caption{Initial tidal and half-mass radius $r_t(0), r_h(0)$ (in natural units) as functions
  of the scaled central potential $W_0$}
\label{tab:rt}
\begin{center}
\begin{tabular}{lll}
$W_0$ & $r_t(0)$&$r_h(0)$\\
\hline
$3$ & 3.15 &0.84\\
$7$ & 7.00 &0.81 \\
$11$ & 6.62&1.01 \\
\hline
\end{tabular}
\end{center}
\end{table}
%%%%%%%%%%%%%%%%%%%%%%%%%%%%%%%%%%%%%%%%%%%%%%%%%%%

\subsubsection{Dependence on $N$}\label{sec:n-dependence}

In our set of simulations we observe a marked dependence of the
dissolution time-scale $t_{dis}$ on the number of particles used (see
Tab.~\ref{tab:runs} and Fig.~\ref{fig:tidal}).  This is not
surprising, as a similar effect is well known in simulations with the tidal force
modeled as in our eq.~(\ref{eq:tidalFORCE}), though using single
particles only: 
the time for the loss of half of
the mass of a cluster is {approximately} proportional to
$t_{rh}(0)^{3/4}$, i.e.
%%%%%%%%
\be
t_{Half} \propto \left(\frac{\ln{(0.11 N)}}{N} \right)^{1/4} t_{rh}(0),
\label{eq:baumgardt}
\ee
%%%%%%%%
(Baumgardt 2001).  Besides empirical support from $N$-body simulations,
this formula can be understood on the basis of the combined effect of
the tidal field and of two body relaxation (Baumgardt 2001).   We find
(Fig.\ref{fig:tidal}) that it also provides a reasonable fit to the
dissolution time $t_{dis}$ of models
with primordial binaries (for the parameters specified in the caption of
the figure), though the observed dependence on $N$ is slightly
stronger.
Therefore we
can take advantage of eq.(\ref{eq:baumgardt}) for the approximate
extrapolation of results {on the dissolution time} $t_{dis}$ to larger numbers of
particles $N$.  Note, however, that eq.(\ref{eq:baumgardt}) is derived in
\citet{bau01} using simulations with $N \leq 16384$; it is expected to
be valid up to $N \lesssim 10^6$ at most, as for larger $N$ the equation
eventually leads to $t_{Half} < \trh0$, a clearly unphysical result.

%{\bf I've altered the next few sentences to look more like
%  hypothetical musings and less like definitive conclusions.}

 Assuming, then, that this scaling applies to the dissolution time
  $t_{dis}$, we can compare our results with those in \fr~ If we
  extrapolate to $N=3~10^5$, we find that $t_{dis}$ decreases by a
  factor $\approx 2.6$ from the results obtained for the simulations
  with $N=4096$ and by a factor $\approx 1.9$ from the runs with
  $N=16384$.  This {would} mean that a simulation starting from
  $W_0=7$ with $10~\%$ binaries and $N=3~10^5$ would take $\approx
  18~t_{rh}(0)$ to dissolve.  For comparison \fr~ measure $t_{dis}
  \gtrsim 38 t_{rh}(0)$\footnote{{\add Note that our definition of
  $t_{rh}(0)$ differs by a factor $\ln(0.11N)/\ln(0.4N)$ from that
  used by \fr. This has been taken into account in this paper to
  ensure a proper comparison.}}  for these initial conditions.  {This
  and other comparisons are given in Table \ref{tab:tbEv} (second and
  fourth columns).}  An understanding of such differences requires
  consideration of the treatment of the tide, and so we defer further
  discussion to Sec.\ref{sec:tidalcut}.
%% Similarly our extrapolation for the $W_0=3$
%% model with $10~\%$ binaries gives $t_{dis} \approx 5.1~t_{rh}(0)$, to be
%% compared with $t_{dis} \approx 14~t_{rh}(0)$ in \fr, and for the
%% $W_0=11$ model with $10~\%$ binaries we estimate a extrapolated $t_{dis}
%% \approx 11.0 ~t_{rh}(0)$ versus $t_{dis} \gtrsim 30~ t_{rh}(0)$ in \fr

%To disentangle the effects of a different tidal field treatment, in
%Sec.\ref{sec:tidalcut} we discuss the comparison with \fr for a series
%of our runs that use a tidal cut-off. 

%%%%%%%%%%%%%%%%%%%%%%%%%%%%%%
\begin{figure}
\resizebox{\hsize}{!}{\includegraphics{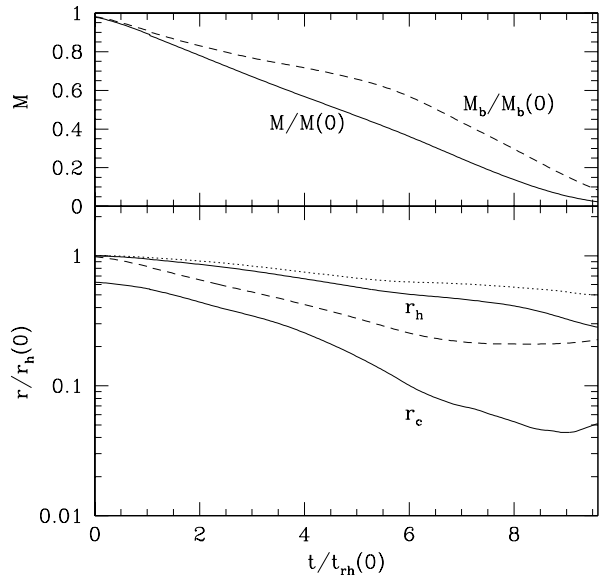}} \caption{Time
dependence of the total mass and of the mass in binaries (upper panel)
and of the half mass and core radius (bottom panel). The dotted line is
the half mass radius for singles while the dashed line in the lower
panel is the half mass
radius for binaries. The lowest curve is the core radius of the system
(in units of the initial half mass radius). The simulation has been
performed with $16384$ particles and $10\%$ of binaries starting from
$W_0=3$. It is the equivalent of Fig~13 in \citet{fre03}}\label{fig:fr13}
\end{figure}
%%%%%%%%%%%%%%%%
  
%%%%%%%%%%%%%%%%%%%%%%%%%%%%%%
\begin{figure}
\resizebox{\hsize}{!}{\includegraphics{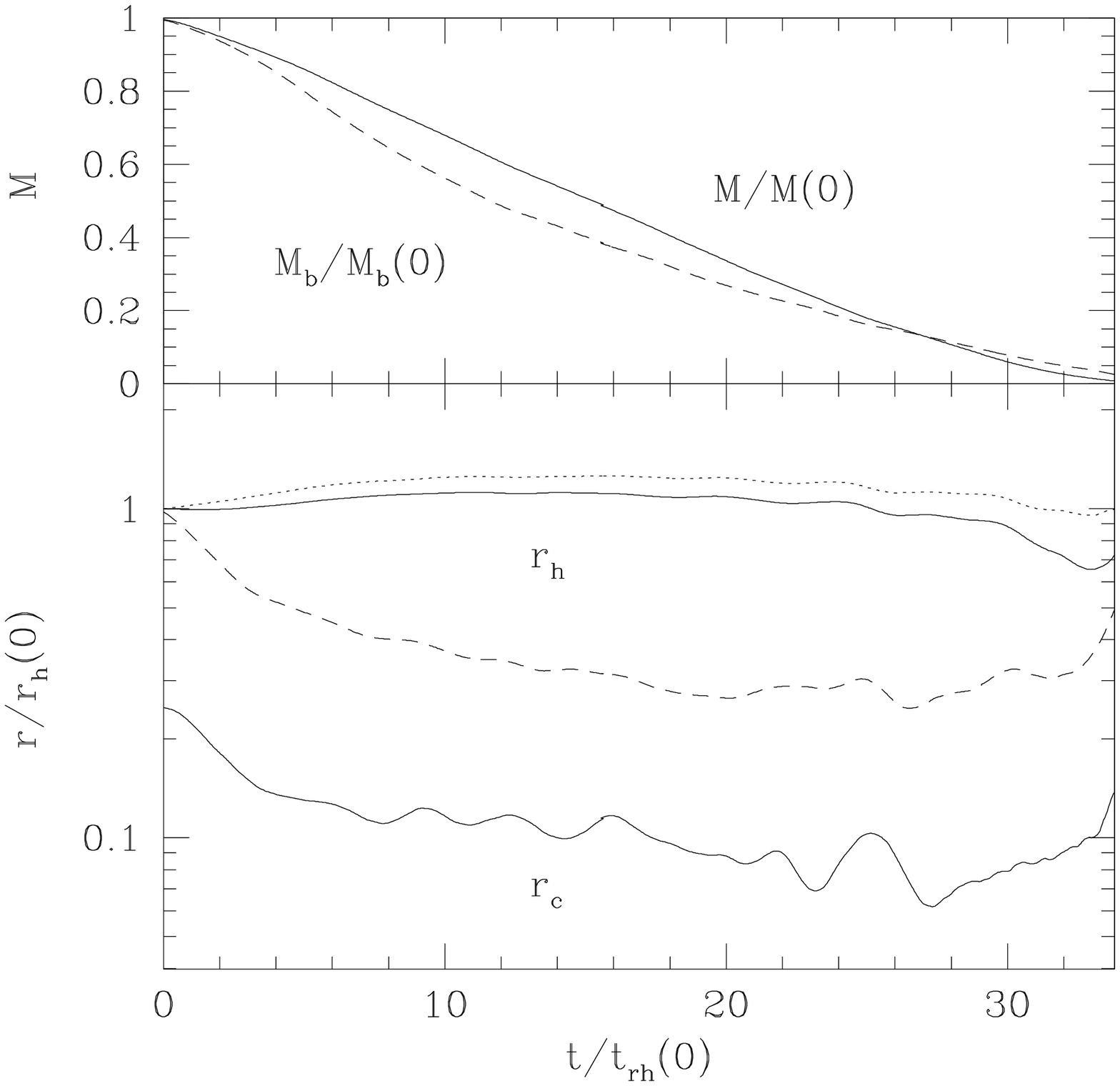}} \caption{Like
 Fig.~\ref{fig:fr13} for a simulation starting from $W_0=7$ with $16384$
 particles and $10\%$ of binaries. It is the equivalent of Fig~10 in
 \citet{fre03}}\label{fig:fr10}
\end{figure}
%%%%%%%%%%%%%%%%

%%%%%%%%%%%%%%%%%%%%%%%%%%%%%%
\begin{figure}
\resizebox{\hsize}{!}{\includegraphics{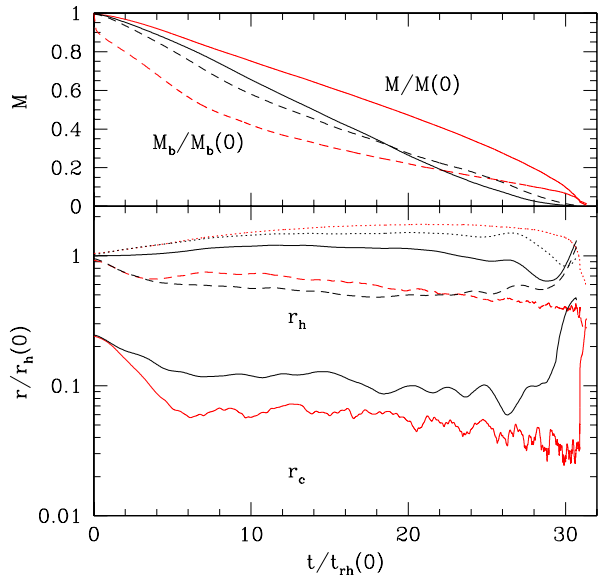}} \caption{Like
 Fig.~\ref{fig:fr13} for a simulation starting from $W_0=7$ with
 $16384$ particles and $20\%$ of binaries. It is the equivalent of
 Fig~11 in \citet{fre03} {~or Fig.3 in \citet{fre05}. The results from a new,
 unpublished run by Fregeau (Fregeau 2006, private communication) are
 overplotted in red.}}\label{fig:fr11}
\end{figure}
%%%%%%%%%%%%%%%%
  
%%%%%%%%%%%%%%%%%%%%%%%%%%%%%%
\begin{figure}
\resizebox{\hsize}{!}{\includegraphics{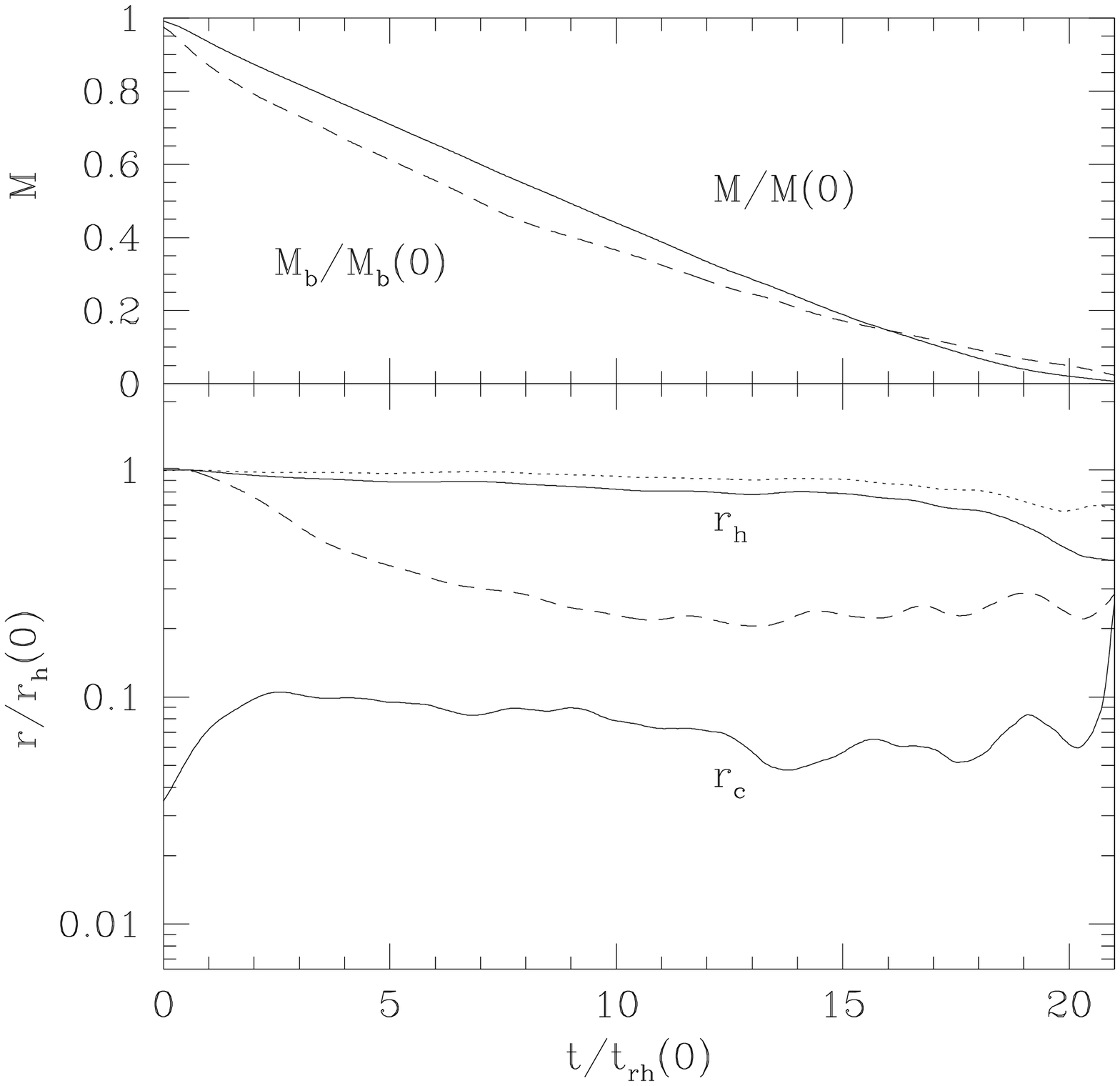}} \caption{Like
 Fig.~\ref{fig:fr13} for a simulation starting from $W_0=11$ with $16384$
 particles and $10\%$ of binaries. It is the equivalent of Fig~12 in
 \citet{fre03}}\label{fig:fr12}
\end{figure}
%%%%%%%%%%%%%%%%

%%%%%%%%%%%%%%%%%%%%%%%%%%%%%%
\begin{figure}
\resizebox{\hsize}{!}{\includegraphics{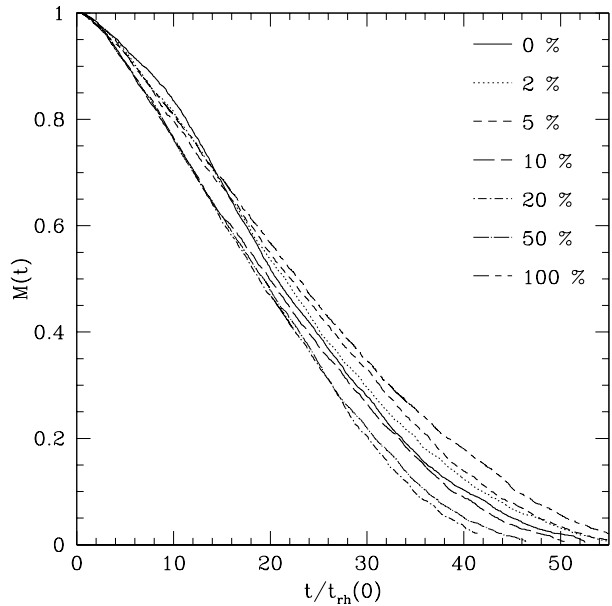}} \caption{Evolution of
the total mass for a King models with $W_0=7$ and different ratios of
primordial binaries.  The simulations have been performed with $4096$
particles.}\label{fig:2}
\end{figure}
%%%%%%%%%%%%%%%%%%%%%%%%%%%%%%

%%%%%%%%%%%%%%%%%%%%%%%%%%%%%%
\begin{figure}
\resizebox{\hsize}{!}{\includegraphics{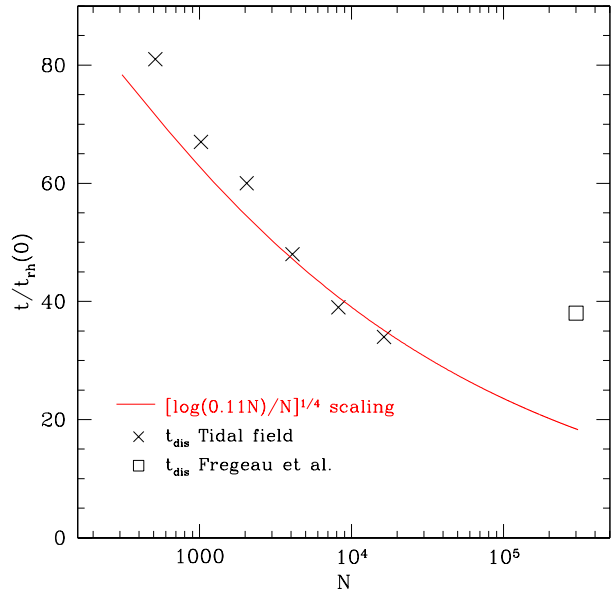}} \caption{Tidal
 dissolution $t_{dis}$ time for models with $W_0=7$, $f=10\%$ and a galactic tidal
 field, for different values of $N$, compared to the result by
 \fr~(square point; obtained using a tidal cut-off). The points
 associated with runs using a tidal field are compared with the theoretical scaling
 (red solid line) given by \citet{bau01} for tidal field runs.
 }\label{fig:tidal}
\end{figure}
%%%%%%%%%%%%%%%%%%%%%%%%%%%%%%

%%%%%%%%%%%%%%%%%%%%%%%%%%%%%%%%%%%%%%%%%%%%%%%%%%%
% table for differences in tidal dissolution time %
\begin{table}
\caption{Comparative results on the dissolution timescale $t_{dis/t_{rh}(0)}$
for simulations with $f=10~\%$}
\label{tab:tbEv}
\begin{center}
\begin{tabular}{l|lll}
$W_0$	& THH field	& THH cutoff & \fr~cutoff \\
\hline
$3$ & $5$ & $15$ & $13$ \\
$7$ & $18$ & $20-40$ & $\gtrsim 38$ \\
$11$ & $11$ & N/A & $27$ \\
 
\hline
\end{tabular}
\end{center}
{Notes: The values for $t_{dis}$ for our runs (THH) have been
extrapolated to $N=3 \cdot 10^5$ from $N=16384$ (see discussion in the
text). In the {second}  column we give $t_{dis}$ for our standard tidal
field runs, extrapolated by using eq.~(\ref{eq:baumgardt}). In the
{third} column we report our results from runs with a tidal cut-off,
with constant extrapolation for $W_0=3$ and a range of values for
$W_0=7$ (see Sec.\ref{sec:tidalcut}). The {fourth} column contains the results from \fr }
\end{table}
%%%%%%%%%%%%%%%%%%%%%%%%%%%%%%%%%%%%%%%%%%%%%%%%%%%

%\subsection{Core and half mass radius: general considerations}
\subsection{Core and half mass radius}\label{sec:rcrhNEW}

\begin{figure}
\resizebox{\hsize}{!}{\includegraphics{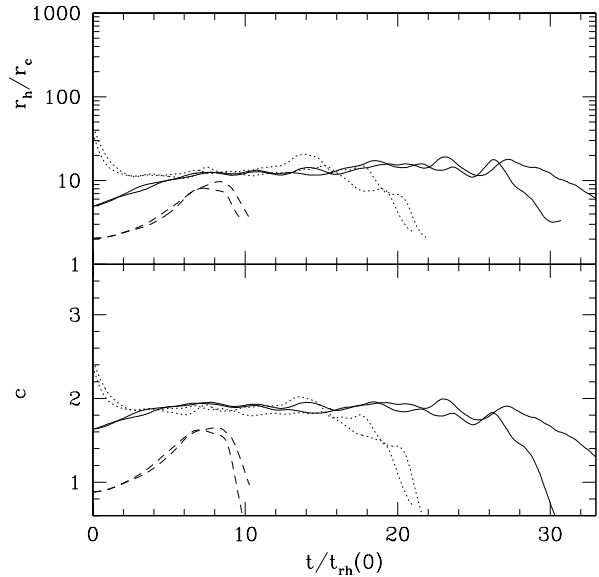}} \caption{Evolution
of the ratio of the half mass to core radius (upper panel) and of the
concentration parameter $c=\log{r_t/r_c}$ for different King models
with $f=10\%,20\%$. The solid line refers to simulations starting from
$W_0=7$, the dotted line to $W_0=11$ and the dashed line to $W_0=3$;
the number of particles used is $16384$.  The evolution quickly erases
differences due to initial conditions, and during binary burning $c$
and $r_h/r_c$ evolve very similarly. This figure can be directly
compared with Fig.~16 in \fr~Note that we kept the same range in the
y-axis to highlight the differences between our runs and
theirs.}\label{fig:fr16}
\end{figure}

%{\bf First we describe the general picture.}
In the presence of a tidal field, the system evolves toward conditions
in the centre very similar to those that we observed for isolated
systems. The core radius evolves, usually in a few $t_{rh}$, 
to reach a ratio $r_c/r_h$ that is close to that
attained at the end of core collapse during the evolution of isolated
Plummer models (see Sec.~4.1 of \hth) with a similar fraction of
primordial binaries.   As is depicted in Fig.~\ref{fig:fr16},
%, which is the equivalent of Fig.~16 in \fr, 
the evolution of the core to half mass radius, after a first
adjustment phase, is largely independent of the initial conditions
considered.  Thereafter this ratio is almost
constant, as long as there are binaries to burn or until tidal
dissolution has reduced the total mass below $10\%$ of the initial
value.  

We discuss the value of the core radius in further detail below, but
summarise here the behaviour of the half mass radius.  This depends on
the strength of the tidal field, which depends on $W_0$
(Sec.\ref{sec:tidal}).  For $W_0 = 7$, $r_h$ remains almost constant,
up to the final stages of the life
of the system, when it eventually falls to zero.  For stronger tidal fields
(models with $W_0=3$ and $11$) the half mass radius decreases
steadily.  Our interpretation is that,  for $W_0 = 7$, the tendency toward
expansion (due to the energy generated in the core) is balanced
by the mass loss due to the tidal field, while the tide
dominates in the other two cases.

\begin{figure}
\resizebox{\hsize}{!}{\includegraphics{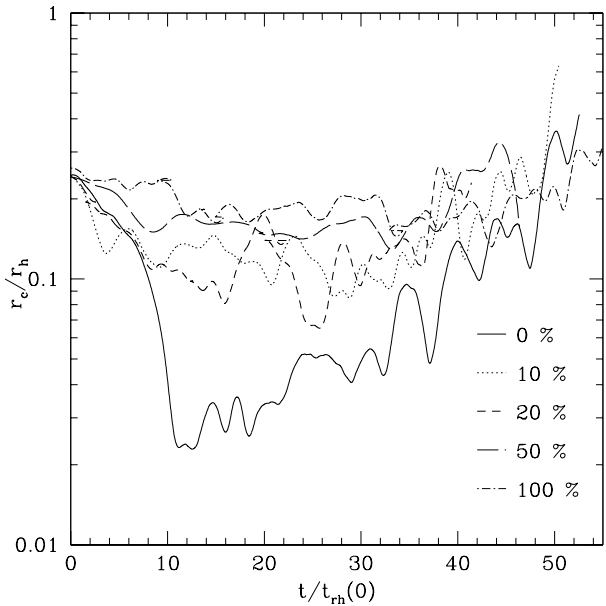}}
\caption{Dependence of the ratio $r_c/r_h$ on time (in units of
$t_{rh}(0)$) 
for simulations starting from a
King model with $W_0=7$ and different ratios of primordial
binaries. The simulations have been performed with $4096$
particles.}\label{fig:3}
\end{figure}

%{\bf The following paragraph incorporates the first paragraphs of 4.3
%  in the previous version. - D}
Now we return to the evolution of the core radius, and consider first the
initial phase of contraction or expansion.
The evolution in the first few relaxation times
depends on how the initial value of $r_c$ compares with the quasi-equilibrium
value attained in the intermediate steady binary burning phase.  Thus
runs starting initially with a relatively big core, such as $W_0=3$
($r_c(0) \approx 0.6 r_h(0)$, Fig.~\ref{fig:fr13}) have a rather deep
collapse; after about $8t_{rh}(0)$ in our simulation with $N=16384$ and $f = 10\%$
 we
measure $r_c/r_h(0)\simeq0.05$.  For $W_0=7$, $f = 10\%$  and $N = 16384$ (see
Fig.~\ref{fig:fr10}) the duration of the initial contraction is
shorter, and thereafter the core radius is somewhat larger.  On the
other hand runs that start with a small, concentrated core ($W_0=11$),
have a first rapid expansion of the core radius (see also
{Fig.~\ref{fig:fr12}}) up to $r_c/r_h(0) \approx 0.09$ for our
$N=16384$ run with $f=10\%$.  This can be understood in terms of the
energy balance of  the core: if the density is
too high, the binary heating exceeds the cooling by heat transport to
the halo, so that the core has to expand to lower density to reduce the
rate of energy generation.  Incidentally, though we have focused on numerical values
for runs with $f=10\%$,  we find that the effect of primordial
binaries saturates by around $f=10\%$ (see Fig.~\ref{fig:3}), much as
we have noted in the evolution of isolated
clusters in \hth~{(see also Vesperini \& Chernoff 1994)}.

\begin{figure}
\resizebox{\hsize}{!}{\includegraphics{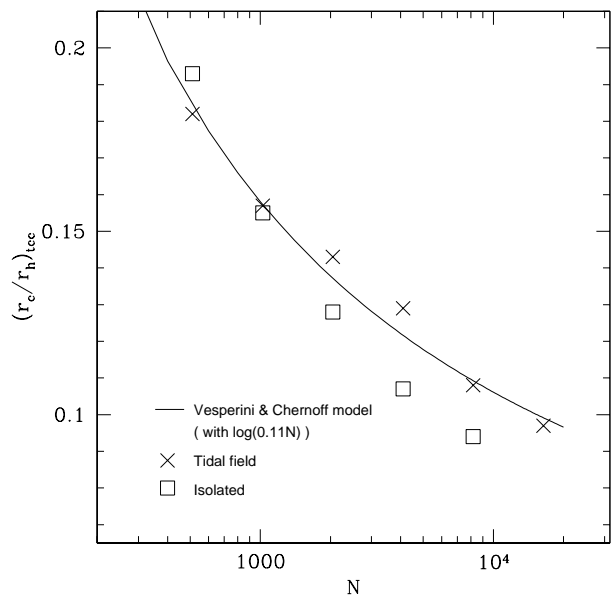}}
\caption{Dependence of the ratio $r_c/r_h$, averaged for
$10~t_{rh}(0)$ after core collapse, on the number of
particles. Simulations start from a King model with $W_0=7$ and
$f=10\%$ using a tidal field (crosses) or isolated runs
(squares). Points associated with $N$ from $512$ to $2048$ have been
obtained by averaging over a sample of $16$ simulations for each
$N$. A comparison with the \citet{ves94} model is drawn: the agreement
(using $log(0.11N)$ instead of $log(0.4N)$) is excellent for the tidal
field runs. The adopted values for the parameters used are given in
the text.}\label{fig:rc_vs_N}
\end{figure}

Now we turn to the value of the core radius during the phase of steady
binary burning, and, in particular, its dependence on the number of
particles.  Our results are summarised in Fig.~\ref{fig:rc_vs_N}. At
variance with \hth, here the \citet{ves94} model provides an excellent
fit to the dependence on $N$ of the ratio of the core to half mass
radius,
provided that in the Coulomb
logarithm the value $0.11$ is used for the coefficient of $N$.  Thus
the formula used for the fit is (see \citealt{ves94} and \hth):
\begin{equation}
\frac{r_c}{r_h} = \frac{\alpha}{\log_{10}(0.11N)}
\frac{\phi_b (1-\phi_b) \mu_{bs} + \phi_b^2 \mu_{bb}}{(1+\phi_b)^4},\label{eq:vc}
\end{equation}
where the various quantities are defined as in \hth. Thus the quantity
$\alpha$ is a parameter depending on $\gamma$ and $v_c/v_h$, where
$\gamma$ is the ratio of the expansion timescale $r_h/\dot r_h$ to
$t_{rh}$, and $v_c$ and $v_h$ are the velocity dispersion in the core
and at the half-mass radius, respectively; we have assumed typical
values for these parameters: $\gamma = 10.5$ and $v_c/v_h =
\sqrt{2}$ \citep{heg92,goo89}. The quantity $\phi_b$ is the binary
fraction in the core defined as $$ \phi_b = \frac{n_b}{n_s+n_b}.  $$
Finally, $\mu_{bs}$ and $\mu_{bb}$ are coefficients for the efficiency
of binary-single and binary-binary burning and depend on the
distribution of binding energy of the binaries. They have been
computed as in ~\hth\ assuming a flat distribution in the logarithm of
the binding energy from $12~kT$ to $300~kT$.{\add $\alpha$, $\mu_{bs}$ and
$\mu_{bb}$ do not significantly depend on the number of particles $N$
(see Sec. 5.1.2 in \hth). }

%%%%%%%%%%%%%%%%%%%%%%%%%%%%%%
%%%%%%%%%%%%%%%%%%%%%%%%%%%%%%

%%%%%%%%%%%%%%%%%%%%%%%%%%%%%%
%%%%%%%%%%%%%%%%%%%%%%%%%%%%%%

%%%%%%%%%%%%%%%%%%%%%%%%%%%%%%
%%%%%%%%%%%%%%%%%%%%%%%%%%%%%%
%%%%%%%%%%%%%%%%%%%%%%%%%%%%%%
\begin{figure}
\resizebox{\hsize}{!}{\includegraphics{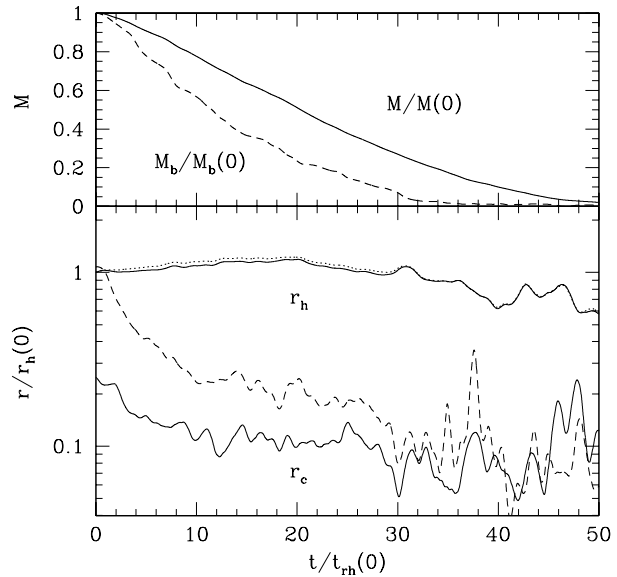}} \caption{Like
 Fig.~\ref{fig:fr13} for a simulation starting from $W_0=7$ with $4096$
 particles and $5\%$ of binaries.}\label{fig:w7_5}
\end{figure}

{With regard to the core radius} there are many quantitative and even
some qualitative differences between our results and the earlier
results of \citet{fre03}.  But they
point out (\citealt{fre05}) that the core radius is one parameter
which has turned out to behave rather differently following the
code improvements summarised in their later paper.  Indeed the one
case for which a direct comparison with their most recent results is
possible is shown in Fig.\ref{fig:fr11} ($W_0=7, f=20\%$).  Both
models (our $N$-body model and their Monte Carlo model) show an
initial contraction followed by a phase of steady binary burning.  The
comparison with the new results by \citet{fre05} is quantitatively
satisfactory, except that, in the phase of steady binary burning, our
core radius is about twice as large as theirs.  However, it is known
that the core radius in this phase is $N$-dependent (see above, and
Paper I), and the sense of the difference is consistent with this.
%Incidentally, it is worth noting that the evolution proceeds much as
%in the $f=10\%$ case.  This {illustrates again the fact that} the
%efficiency of the binaries in the core saturates when
%$f\gtorder10\%$.% (see Paper I and \citet{ves94}).

As already mentioned in Sec.\ref{sec:picture}, the {\sl late} evolution of
the core radius depends on whether the binary population can be
heavily depleted by the time of dissolution, $t_{dis}$.  The evolution of the
binary population is discussed in the following section, and so we
defer further discussion until then.

\subsection{Evolution of the primordial binary population}\label{sec:binpop}

%%%%%%%%%%%%%%moved here from Sec.3

\begin{figure}
\resizebox{\hsize}{!}{\includegraphics{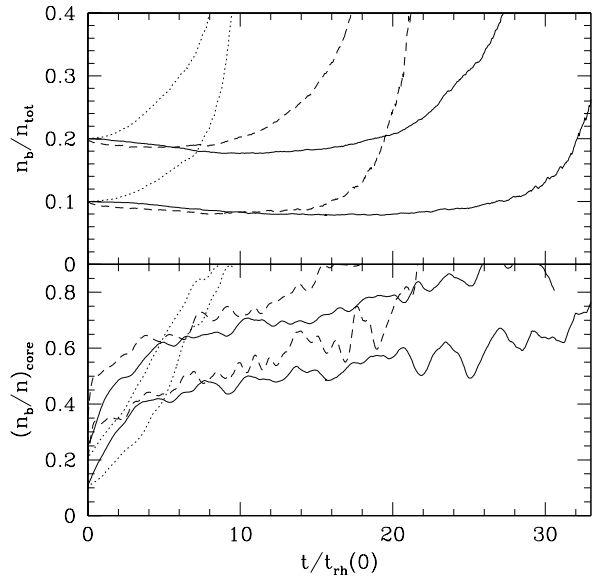}}
\caption{Dependence of the binary fraction on time (in units of
$t_{rh}(0)$) 
for simulations starting from King models
with $W_0=3$ (dotted lines), $W_0=7$ (solid lines), and $W_0=11$ (dashed
lines) for $f=10\%,20\%$. The upper panel refers to the total number of
binaries, the lower to the binaries in the core. The simulations have
been performed with $16384$ particles.}\label{fig:10}
\end{figure}

\begin{figure}
\resizebox{\hsize}{!}{\includegraphics{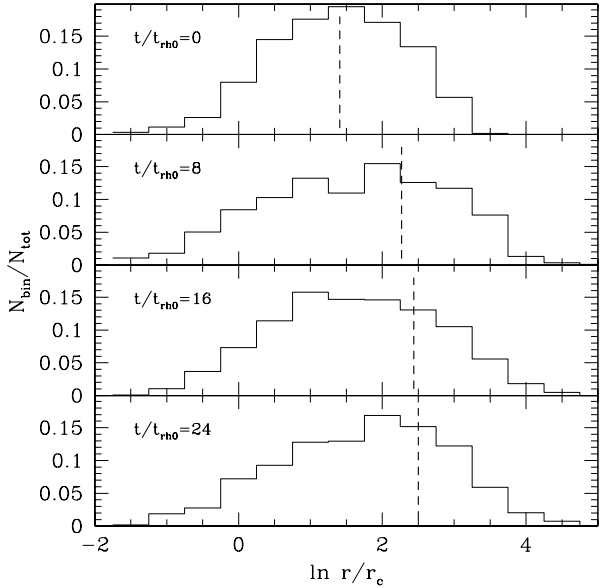}} \caption{Evolution of
the distribution of binary radius in units of the instantaneous
core radius. The ordinate is the number of binaries per bin, which are
of length $0.5$  in $\log_e$. Curves have unit normalization, and are
shown for $t/t_{rh}(0) = 0, 8, 16, 24$. The dashed line is the position
of the half mass radius. The simulation has been performed with $16384$
particles and $20\%$ of binaries starting from a King model with
$W_0=7$.}\label{fig:rbin}
\end{figure}

The destruction of the primordial binary population is depicted in the
top panel of Figs.~\ref{fig:fr13}-\ref{fig:fr12}.  The number of
binaries decreases due to both destruction of the softer pairs in the
centre of the system and the ejection of stars across the tidal
boundary. 
The changing relative contribution of these two processes can be
inferred from Fig.~\ref{fig:10}, upper panel.  At first the ratio of
binaries to singles decreases, as at the beginning of the simulation
relatively many soft binaries (i.e. those with $E_b \lesssim 15kT$)  are
present. However at later times the relative number of binaries 
to singles starts to rise, as most of the surviving binaries
are hard to disrupt in three- or four-body encounters. {In addition tidal
ejection is more probable for singles, as binaries have sunk toward
the core of the cluster by mass segregation (see bottom panel in
Figs.~\ref{fig:fr13}-\ref{fig:fr12} and Fig.~\ref{fig:rbin}).}

There are some quantitative differences between our results and those
of \fr, even after the recent improvements to their technique.  For
example they point out (\citealt{fre05}) that, for a model with
$W_0=7$ and $f=20\%$, the binary fraction shrinks to about $10\%$
before recovering to about its initial value shortly before $t_{dis}$.  This differs considerably from our result in
Fig.\ref{fig:10} (upper panel).  Another example is illustrated in the
upper panel of Fig.\ref{fig:fr11}, where it is clear that the
disruption of binaries in the most recent available run by \fr~
appears to be rather faster than in our run starting from similar
initial conditions.

Now we consider models in which the population of binaries can be
strongly depleted before dissolution of the cluster, i.e. $t_{bb}<t_{dis}$.
We have already remarked (Sec.\ref{sec:tidal}) that $t_{dis}$
 increases as $N$ decreases, and is longest for $W_0 = 7$
(out of the three values we adopted).  It has also been argued
(Sec.\ref{sec:picture}) that the time taken to burn most binaries is
roughly proportional to $f$.
Based on these considerations, it may be expected that the only runs
that may
burn most of their binaries within the lifetime of the system are those in
which $f$ or $N$ is
sufficiently small, and that intermediate values of $W_0$ are most
favourable.   Indeed, we found that 
the only runs which did so were those with $W_0=7$ and $N \lesssim 4096$ and $f
\lesssim 5\%$.   (Note, however, that $W_0 = 7$ is the only value of
$W_0$ for which we carried out runs with $N<4096$, see Table
\ref{tab:tbEv}.)  At the point where the binaries become heavily
depleted, there is a deep core contraction,
similar to what happens in runs with single stars only; however, here $N$
is too low for gravothermal oscillations to occur.  In
Fig~\ref{fig:w7_5} we depict the evolution of a $W_0=7$ model starting
from $f=5\%$ and $N = 4096$: by about  $ 30~t_{rh}(0)$ the fraction of binaries
has fallen below $20~\%$ of its initial value and the core radius contracts
sharply.  

%Let us consider systems described initially by a King model, with a
%self-consistent tidal field.  We find that  dissolution happens on a timescale below $50
%\trh0$ if
%$N\gtorder4096$, for every value of the scaled central potential  $W_0$.  Provided that the number of particles is not
%unrealistically low, therefore, tidal dissolution is (much) faster than
%the depletion of a $10~\%$ primordial binary population due to three-
%and four-body encounters.   Thus gravothermal oscillations are not
%observed, and after the initial adjustment of the core, 
%the evolution of the core to half-mass
%radius  proceeds very similarly to the phase
%of steady binary burning which occurs 
%where no tidal field is present (when escape is much reduced).  
%By contrast we shall see (Sec.\ref{sec:rcrhNEW}) that runs starting from
%similar initial conditions but with less than $5~\%$ primordial
%binaries may undergo a deep core contraction after depleting the
%binaries and before being tidally dissolved.  Similarly, as dissolution
%takes longer for smaller $N$, the binary population may be almost
%completely destroyed before dissolution, though in this case no
%gravothermal oscillations are observed.
%%%%%%%%%%%%moved here from Sec.3

%{\bf MT: two paragraphs erased and summarized at the beginning of the section}

Next we comment in particular on the evolution of the binary population
in the core.  At the end of our simulations starting from $f$ {$ =
10-20\%$} we find that the fraction of binaries in the core has
increased by at least a factor $4$. In general the density of binaries
in the core is in broad agreement with the numbers observed in
\hth~and does not depend significantly on the details of the initial
conditions if $f$ is in this range (see bottom panel of
Fig.~\ref{fig:10}, where runs starting from $W_0=7$ and $W_0=11$ show
a rather similar behaviour).  However there are differences in detail
in the time evolution.  In an isolated
cluster the proportion of binaries in the core, after the sharp
initial increase caused by mass segregation, declines slowly; in these
tidally limited models, on the other hand,
there is a slow increase.  The $W_0=3$ King model is an extreme
example, as there is an enhanced rate of ejection of singles relative
to the rate of disruption and ejection of binaries, because of the
relatively small tidal radius; in this situation the binary fraction
in the core increases rapidly throughout the evolution.

\begin{figure}
\resizebox{\hsize}{!}{\includegraphics{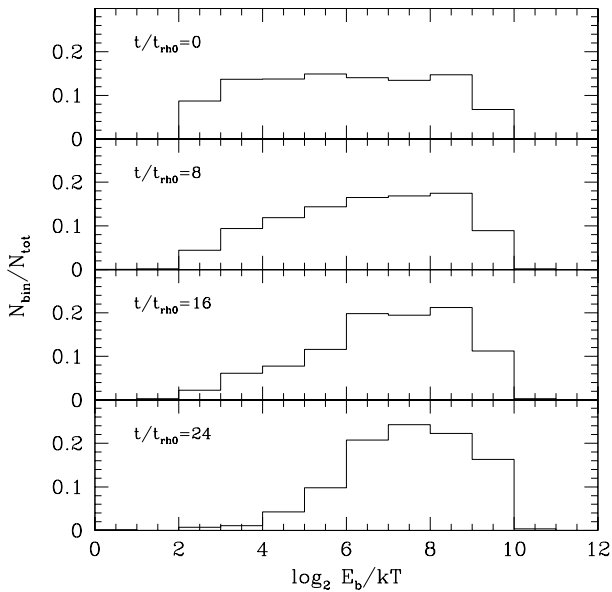}}
\caption{Evolution of the distribution of binary binding energy (given
in units of $kT$).  The ordinate is the number of binaries per bin,
which are of unit length in $\log_2$. Curves have unit normalization,
and are shown for $t/t_{rh}(0) = 0, 8, 16, 24$. The simulation has
been performed with $16384$ particles and $20\%$ of binaries starting
from a King model with $W_0=7$.}\label{fig:ebin}
\end{figure}

\begin{figure}
\resizebox{\hsize}{!}{\includegraphics{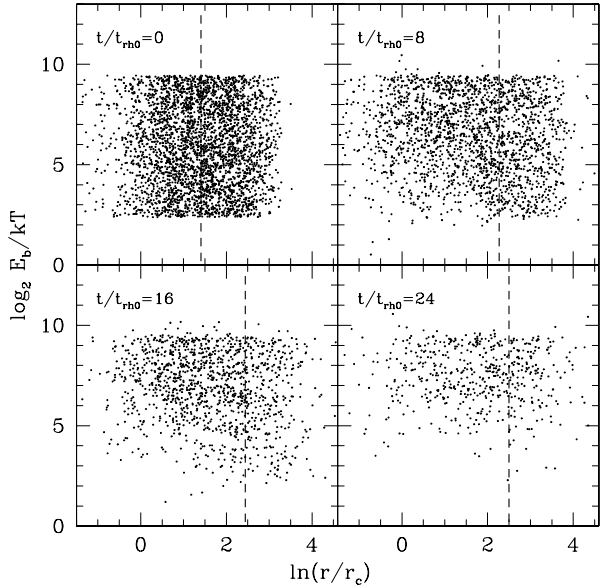}}
\caption{Evolution of the distribution of binaries in  radius
and binding energy.  The radius is given in units of the
instantaneous core radius; the energy in units of $kT$. The panels
refer to $t/t_{rh}(0) = 0, 8, 16, 24$. The dashed line is the position
of the half mass radius. The simulation has been performed with
$16384$ particles and $20\%$ of binaries starting from a King model
with $W_0=7$. It can be compared to Fig.~15 in \citet{fre03}}\label{fig:binplot}
\end{figure}

The evolution of the internal binding energy (see Fig.~\ref{fig:ebin})
is similar to that observed for isolated systems, with a preferential
decrease in the number of less bound binaries.
% with the evolution
%going toward a self similar profile \citep{goo93}{\bf Does it?  And I
%don't understand the next point!:}. However, due to the shorter
%time-span of the run (because of tidal evaporation), the self similar
%profile in $N(E)$ is not reached in our simulation with $N=16384$,
%$W_0=7$ and $f=0.2$ (however that happens for $N=4096$, $W_0=7$ and
%$f=0.1$).
{Furthermore,}
as expected, there is a {significant} correlation of the internal binding
energy with the radius (see Fig.~\ref{fig:binplot}). After
approximately $15~\trh0$ the survival probability in the core for
binaries with binding energy below $\approx 20~kT$ is low, as they
have either been destroyed or hardened.  We thus see that softer binaries
are mainly present around the half mass radius and in the halo, while
the core is dominated by hard binaries (with occasionally some short-lived
low energy binaries that have been formed dynamically). At later
stages of the evolution (see bottom right panel of
Fig.~\ref{fig:binplot}), the softer binaries have almost completely
disappeared.

This picture is in qualitative agreement with the results already
known in the literature (see e.g., \citealt{mcm94,gie00} and \fr).  As
our resolution is limited by the modest number of binaries that are
present in our runs, it is hard to attempt any quantitative
comparison.
% with Fig.~15 {in the earlier paper} of \fr~ However, it
%seems that the depletion of low binding energy binaries is more
%efficient in their simulations: between $10$ and $20$ $t_{rh}(0)$ they
%observe a complete disruption of binaries located almost up to {the}
%half mass radius with energy in the range up to $\approx
%30~kT$. (Please note that their definition of $kT$ differs from ours,
%as they define $T$ as the mean kinetic temperature of stars in the
%\emph{core}). In our runs even at $t=24~t_{rh}(0)$ we observe the
%presence of a few binaries with $E_b \lesssim 15~kT$ located in the
%core. 

%%%%%%%%%%%%%%%%%%%%%%%%%%%%%%
%%%%%%%%%%%%%%%%%%%%%%%%%%%%%%

%%%%%%%%%%%%%%%%%%
%%%%%%%%%%%%%%

%%%%%%%%%%%%%%%%%%
%%%%%%%%%%%%%%

%%%%%%%%%%%%%%%%%%
%%%%%%%%%%%%%%

\section{Tidal cutoff runs: Results and Comparison}\label{sec:tidalcut}

It was pointed out in Sec.\ref{sec:n-dependence} that our result on
the lifetime differed significantly (at least in the one case
considered) from that of \fr~ This conclusion, however, required a
knowledge of the $N$-dependence of the lifetime, and it is known that
this in turn depends on the treatment of the tide, which differs
between the work of \fr~ and the $N$-body simulations which we have
presented so far.  For the case of a tidal cutoff, as adopted by \fr,
it was found by \citet{bau01} that the half-mass dissolution
time-scale $t_{Half}$ (expressed
in relaxation times) does not depend on the number of particles of the
system (as we assumed for the purpose of our comparison) if $N \gtrsim
4096$.  Note, however, that \citet{bau01} used a \emph{fixed} tidal
radius and, as we have mentioned, measured the timescale of mass-loss
by the half-mass time $t_{Half}$, whereas our discussion has focused on the time
for loss of 98\% of the mass, $t_{dis}$.

Because of these complications, 
for a better comparison with \fr we have run a second series of
simulations, in which 
the tidal field we
have used hitherto has been replaced by a tidal cutoff.  We have 
ensured that the cutoff radius decreases in the correct manner
(eq.(\ref{eq:rt})) as the
mass inside this radius decreases, exactly as adopted by \fr  The
initial parameters of these  models  are $W_0=7$ and $W_0=3$,
$f=10\%$, with  $N$ in the range
$512$ to $16384$.  

For the series starting from $W_0=3$ with $f=10~\%$ (i.e. with a
strong tidal field, Sec.\ref{sec:w-dependence}), we do not observe a
significant $N$-dependence of the half-mass dissolution time
$t_{Half}$ in units of the relaxation time,{\add except for a modest
($\approx 10\%$) variation from $N=4096$ to $N=8192$.}  This is
consistent with the experiments (without primordial binaries) by
\citet{bau01}, despite his use of a constant tidal field.  Moreover,
it is also consistent with unpublished results by one of us using the
initial conditions of the collaborative experiment summarised in
\citet{heg03}, except for replacement of the tidal field by a tidal
cutoff; in this case the cutoff was evolved in the correct manner as
mass was lost.  For the $W_0=3$ runs with tidal cut-off the agreement
with the dissolution time $t_{dis}$ of \citet{fre03} is good: we
extrapolate $t_{dis} \approx 15~t_{rh}(0)$ and they have
$t_{dis}=13~t_{rh}(0)$.
%{This agreement remains tentative, however,
%until the effects of recent improvements in their code are known. }

\begin{figure}
\resizebox{\hsize}{!}{\includegraphics{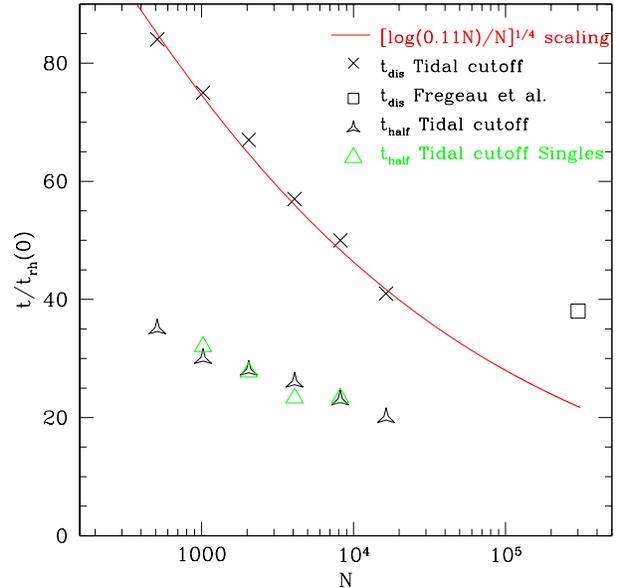}} \caption{Tidal
 dissolution time $t_{dis}$ for a $W_0=7$ $f=10\%$ model with a tidal
 cutoff, with different values of $N$, compared to the result by
 \fr~(square point; obtained using a tidal cut-off). The lower series
 (3-sided stars) gives the half mass dissolution time $t_{Half}$,
 compared to some runs starting from $W_0=7$ and no binaries
 (triangles). {\add The scaling with $N$ of both $t_{dis}$ and $t_{Half}$
 appears similar.}}\label{fig:tidalCUT}
\end{figure}

By contrast, for the series with $W_0=7$ and $f=10~\%$ {and a cutoff}, we
observe an $N$-dependence similar to that found when  a tidal field is
used, as in our
standard runs (see Fig.~\ref{fig:tidalCUT}).   On the other hand it is not
known why the $N$-dependence should differ for the cases $W_0 = 3$ and
$7$ when a cutoff is used.  One possibility is that it is 
a ``small-$N$'' effect which is more important in a system with a
small core; if so, the $N$-dependence of the dissolution time $t_{dis}$ will eventually
flatten out for larger  $N$.  The presence of primordial binaries does
not seem to be the issue; we have run a few
simulations with a tidal cut-off, starting from a $W_0=7$ model  and no
binaries, and find that the half-mass dissolution time $t_{Half}$ is almost equal to that
measured when binaries are present.  
%Most
%probably the difference between \citet{bau01} and our runs lies in the
%fact that \citet{bau01} kept the cutoff radius constant for the whole
%simulations, while we follow \fr~in updating it as the system loses
%mass.
%%% COMMENTED OUT AFTER THE NEW RUNS mt15/06/2006 %%%
%One possible explanation along this line could
%lie in the fact that the relative core density of binaries with
%respect to singles in the steady binary burning phase increases with
%$N$ (see also Fig.~21 in \hth). In that case for runs with higher N it
%would be more probable for single stars to leave the system, so that
%the total mass (and the tidal radius) would decrease faster. This
%scenario would not apply to $W_0=3$ as the system loses more than half
%of the initial mass before the end of the core collapse. To test this
%explanation we have run a few simulations starting from a $W_0=7$
%model with tidal cut-off and no binaries.
\citet{tf05}
have also observed differences (in the scaling of the half-mass time
$t_{Half}$ with $N$) between models with $W_0=3$ and $7$, though their
models adopted a tidal field, and the potential was softened.  They
tentatively attribute the difference to the time taken for a particle
to escape.

Whatever the explanation, this (unexpected) $N$ dependence of $t_{dis}$
 in our tidal cutoff runs {with $W_0=7$} makes the
extrapolation of our results obtained with the tidal cut-off
prescription highly uncertain.  {While the dissolution time of our
model with $W_0=7,f=10\%$ and $N=16384$ in a tidal cutoff is
$t_{dis}\simeq41.2t_{rh}(0)$ (Table \ref{tab:runs}),} the inferred dissolution time for a
$W_0=7$ $f=10\%$ model with $N=3 \cdot 10^5$ may be in the range of
$20-40 t_{rh}(0)$ {(Table \ref{tab:tbEv}, col.3)}, { depending on
whether $t_{dis}$ is proportional to $t_{rh}(0)$ or scales
in accordance with eq.(\ref{eq:baumgardt})}. If the actual value is
close to the upper limit, then the tidal dissolution time $t_{dis}$ in our
$W_0=7$ cutoff models would compare marginally well with \fr, as they
find $t_{dis} \gtrsim 38$.  {Furthermore, recent improvements to their
code appear to have reduced the lifetime slightly for a model with
$W_0=7,f=20\%$, and a similar reduction for $f=10\%$ would improve the
agreement between our $N$-body and their Monte Carlo results.}

%We stress, however, that a physically correct
%treatment of the tidal force requires the use of a field and not a
%cutoff, so that a realistic extrapolation to $N=3 \cdot 10~5$ has to
%be done within our standard tidal field treatment described in
%Sec.~\ref{sec:tidal}.

{ We close this section with a brief comparison between other
  properties of our runs with a tidal field and those with a tidal
  cutoff.  Because of the limited scope of our runs with a tidal
  cutoff, we are concerned less here with general trends; but these
  results will be useful for comparison with simplified treatments in
  which it is necessary to use a cutoff, such as Fokker-Planck models;
  they also illustrate the kinds of results which may be sensitive to
  the manner in which  the tidal boundary conditions are 
  treated.}

\begin{figure}
\resizebox{\hsize}{!}{\includegraphics{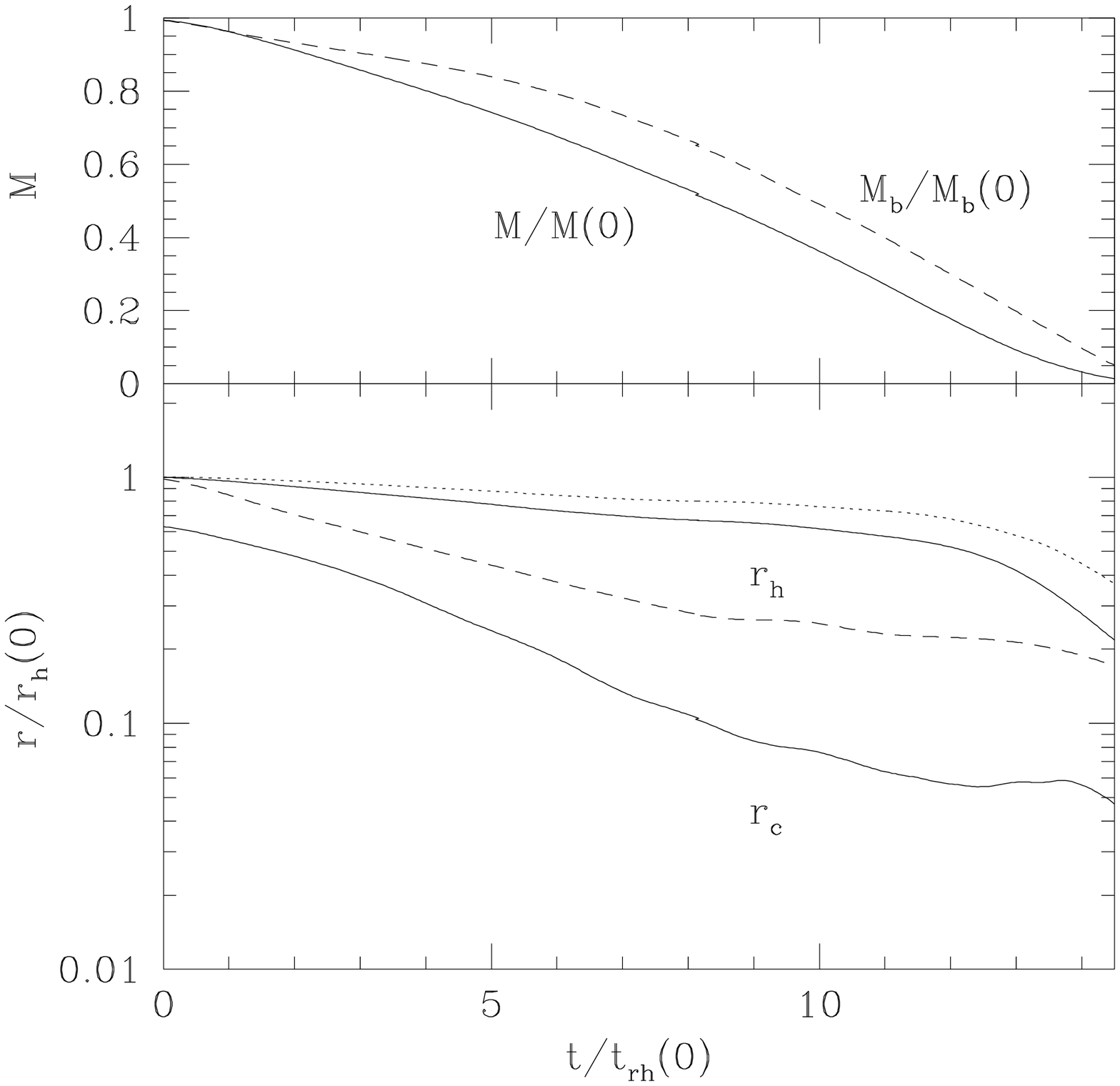}} \caption{Like
 Fig.~\ref{fig:fr13}, but for a tidal cutoff simulation starting from
 $W_0=3$ with $16384$ particles and $10\%$ of binaries. It is the
 equivalent of Fig~13 in \citet{fre03}}\label{fig:fr13CUT}
\end{figure}

    For our run with $W_0=3$,
  $f=10\%$, $N=16384$ and a cutoff (Fig.~\ref{fig:fr13CUT}), the initial
  contraction of the core radius is slower than that observed in the
  run with a 
  tidal field (see Fig.~\ref{fig:fr13}): after $8t_{rh}(0)$, $r_c$
  is approximately twice as large in the cutoff run.  However just before tidal
  dissolution the values of the core radius  in the two
  runs are quite similar.  Actually, if the time is normalized by the
  dissolution time $t_{dis}$, 
  the evolution of $r_c$ looks very similar, except at the very last
  moment, where $r_c$ decreases for the run with a cutoff.
% This may however well be due to the noise present
%where only few bound particles remains in the simulations. 
%%%%{\bf I've deleted a comparison with the old Fregeau et al data here
%%%%  and at the end of the following paragraph. - D.}
%% The comparison with the run by \fr~ is better in this case, but still
%% we have a significant quantitative difference: after $9~t_{rh}(0)$ we
%% have $r_c/r_h(0) \approx 0.1$, while \fr~ report $r_c/r_h(0) \approx
%% 0.2$. This difference is expected to increase after extrapolation of
%% our results to number of particles they used.

\begin{figure}
\resizebox{\hsize}{!}{\includegraphics{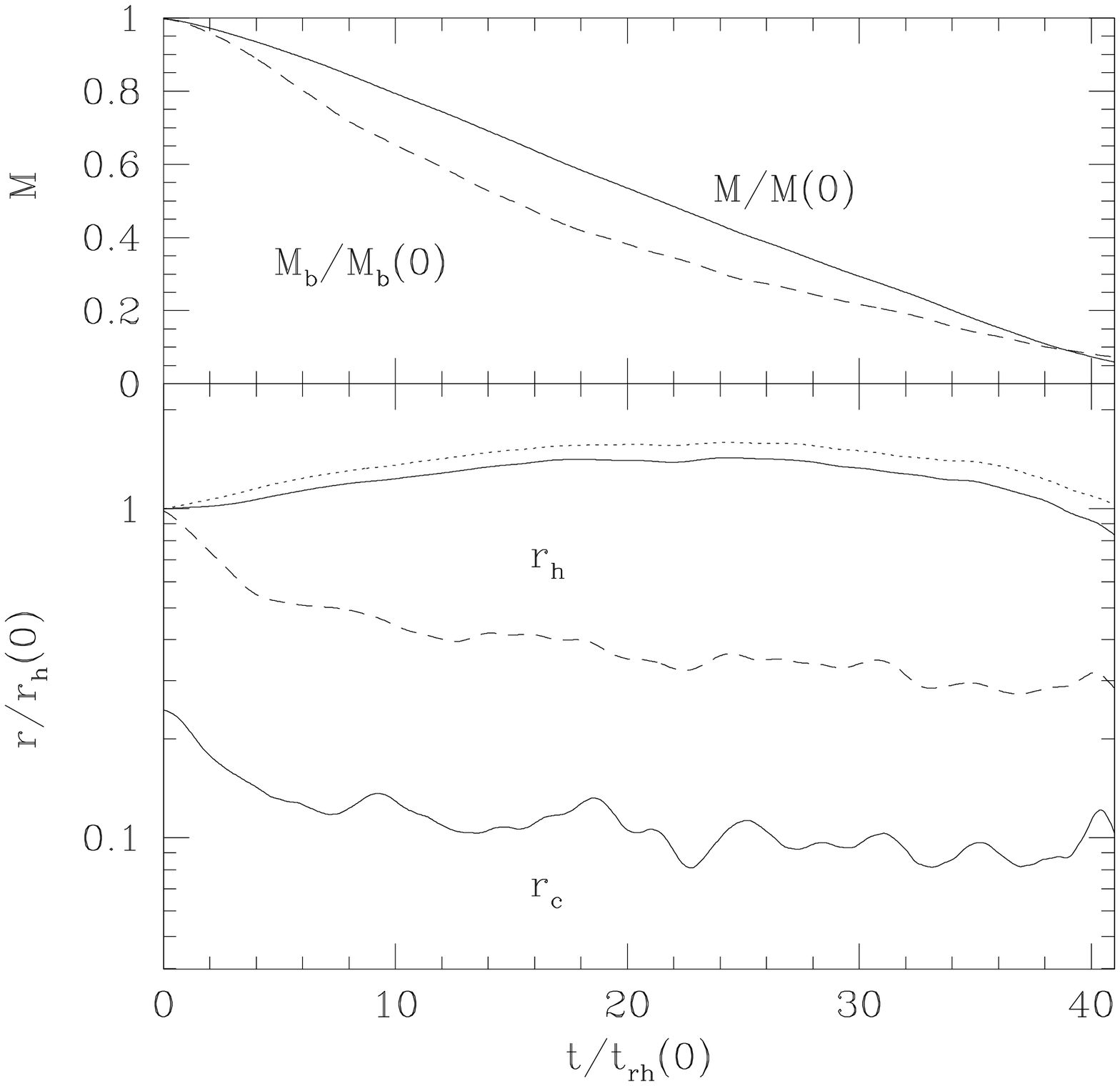}} \caption{Like
 Fig.~\ref{fig:fr10} for a tidal cutoff simulation starting from
 $W_0=7$ with $16384$ particles and $10\%$ of binaries. It is the
 equivalent of Fig~10 in \citet{fre03}}\label{fig:fr10CUT}
\end{figure}

By contrast, the run with $W_0=7$, $f=10~\%$, $N=16384$ and a cutoff (Fig.~\ref{fig:fr10CUT}) is
much more similar to the corresponding run with a tidal field 
(Fig.~\ref{fig:fr10}).   The dissolution time $t_{dis}$
scales for the two tidal treatments differ by only $\approx 25\%$, and the
core radius evolves in a similar way.  However the half mass
radius for the run with a cutoff initially increases in much the same
way as in the run reported by
\fr, while in our run with a tidal field it remains almost constant. 
%The evolution of the concentration parameter $c=log(r_t/r_c)$ for both
%$W_0=3$ and $W_0=7$ and a cutoff is similar to that observed in the
%runs with a tidal field;
%in particular, $c$ is still decreasing at the end of the simulation.
%(while \fr~ report an increase of $c$).   

%The evolution of the fraction of binaries in our runs with a cutoff is
%similar to that observed for the runs with a field, and at the end of the
%simulation we always have a value higher than the initial value.  At
%any given time, however,  the binary fraction is higher for runs with
%a tidal field, where the tidal dissolution ({which gives preference to} singles over
%binaries) acts on a faster timescale. 

%To conclude we present in Fig.\ref{fig:newsim} the results of a
%simulation starting from a $W_0=7$ model with $f=20\%$ performed by
%\fr~using the improved version of their code \citep{fr05} and compared
%with our run for the same initial conditions (but with N=16384, and
%tidal cutoff). The agreement is XXXXXXX

%%%%%%%%%%%%%%%%%%%%%%%%%%%%%%
%%%%%%%%%%%%%%%%%%%%%%%%%%%%%%
  
%%%%%%%%%%%%%%%%%%%%%%%%%%%%%%
%%%%%%%%%%%%%%%%%%%%%%%%%%%%%%%
%%%%%%%%%%%%%%%%%%%%%%%%%%%%%%%
%%%%%%%%%%%%%%%%%%%%%%%%%%%%%%%

\section{Discussion and Conclusions}\label{sec:con}

In this paper we have continued our investigation of the dynamical
evolution of stellar systems with primordial binaries. In \hth~ we
considered simple isolated models with equal mass stars and a primordial
binary population in the range $0-100~\%$.  Then the main
parameters that control the evolution of the system are the primordial
binary fraction $f$ and the number of bodies $N$. However, for $f \gtrsim 10\%$ the efficiency of
energy generation due to three- and four-body encounters saturates, so that
little difference was observed in the size of the core radius for runs
starting with the same number of particles and $f \gtrsim 10\%$. This
behaviour had already been predicted by the theoretical model of \citet{ves94},
and  our runs confirmed this expectation. Interestingly, however, the
scaling of the core radius with the number of particles was not quite
the same as that
predicted by \citet{ves94}.

Here we have added an important new ingredient to our simulations: a
tidal field. The presence of a tidal field introduces a new time-scale
in the simulation, the time needed for the dissolution of the system,
which is usually shorter than the time needed to deplete the
primordial binary population for simulations starting from $f \gtrsim
10\%$ and $N\gtrsim4096$. We have considered runs starting from King
models with initial central potential $W_0=3,7,11$ (and thus a
different intensity of the tidal field, in the units adopted) with $0$
to $100~\%$ primordial binaries, and $N$ from $512$ to $16384$.

By comparison with the results of \hth, our Fig.~\ref{fig:fr16} clearly
confirms that the evolution of the core to half mass radius proceeds
much as for isolated models, after an initial transient; the models
starting with $W_0 = 7$ and $11$ evolve very similarly, for initial
binary fractions of $f=10\%,20\%$. The King model with $W_0=3$,
however, nicely illustrates the effects of a strong tidal field: the
evolution of $r_c/r_h$ tries to go toward the steady value reached by
the other runs with higher $W_0$, but the tidal field manages to
dissolve the system just when this common $r_c/r_h$ value is about to
be attained.

One  objective of this paper was also to compare the results
obtained by means of our direct $N$-body simulations with the outcome of
simulations able to employ realistic number of particles by using
approximate methods, like Monte-Carlo and/or Fokker Planck
approaches. Just as in \hth~ we compared our work with the milestone
{study} of \citet{gao91}, here we took as comparison the recent work
carried out by \fr~ {There are considerable differences with the
results reported in \citet{fre03}, but they have since reported
that several results  change significantly following an improved treatment of binary interactions
\citep{fre05}.  As a result of these improvements they have discovered
that the emission of energy in encounters with binaries was too high
in the earlier models.  Therefore we have focused our comparison on
one case which they illustrated in some detail in their later paper.}
While it is promising to report that most differences have been
cleared up by the improvements, there are exceptions.  One is that the
binaries still appear to be depleted too rapidly, even in the most
recent runs (Fig.\ref{fig:10}; cf. \citet{fre05}, Fig.3). There remain
uncertainties also about the lifetime, but this depends on perplexing
problems in the scaling with $N$: we find evidence that the tidal
cutoff scaling derived by \citet{bau01} for single stars with a fixed
cutoff radius does not seem to be applicable to the self-consistent
prescription of a cutoff radius prescription used by us (for some
simulations) and by \fr~ Therefore the best
approach to understanding the remaining differences between our runs and those of
\fr~ would be to directly compare a set of Monte Carlo simulations starting from
the same initial conditions as we have adopted.

We finally note that, in the presence of a tidal field, the prediction
of the \citet{ves94} model, on the size of the core as a function of
the number of particles used in the simulation, is in detailed
quantitative agreement with the values measured in our set of
simulations with $W_0=7$. This fact comes as a surprise, because, as
we discussed in \hth, the $N$ dependence given by \citet{ves94} was
not verified for isolated clusters: however, the model does not make
any assumption about the presence or absence of a tidal field. The
presence of a tidal field seems thus to introduce a difference in the
scaling with $N$ of the ratio $r_c/r_h$ in the steady burning phase,
which it would be interesting to understand better.

%In the next paper of the series we will continue to add physical
%ingredients to our simulations by considering the effect of an
%intermediate massive black hole on the evolution of star clusters with
%primordial binaries.

\section*{Acknowledgements}

We are indebted to Sverre Aarseth for providing his code NBODY6. {
We thank John Fregeau, Fred Rasio and Atakan G\"urkan for interesting
discussions and for critical comments on a draft of the manuscript;
John Fregeau also kindly supplied us with some of his unpublished data. We are very grateful to the referee for his help in making us see
places where improvements were needed, and in particular for his
observation that the half-mass time scales well with the total-mass
relaxation time.  Finally we thank Andreas Ernst for useful
suggestions.}
MT thanks Prof. Mineshige for his kind hospitality at the Yukawa
Institute at Kyoto University, through the Grant-in-Aid 14079205 of
the Ministry of Education, Culture, Sports, Science and Technology,
Japan. P.H. thanks Prof. Ninomiya for his kind hospitality at the
Yukawa Institute at Kyoto University, through the Grants-in-Aid for
Scientific Research on Priority Areas, number 763, ``Dynamics of
Strings and Fields", from the Ministry of Education, Culture, Sports,
Science and Technology, Japan.

\bsp

\label{lastpage}

%\newpage 
%.
%\newpage

%%%%%%%%%%%%%%%%%%%%%%%%%%%%%%
%     TABLE WITH OUR RUNS    %
%%%%%%%%%%%%%%%%%%%%%
\begin{center}
\begin{table*}
\caption{$N$-body models with \fr-like initial conditions. In this
table we report summary information for all runs with $N \geqslant
4096$. For runs with smaller $N$ we report averaged values over $16$
simulations for each entry line. The entries at the bottom of the
table refer to simulations with tidal cut-off as in \fr~Column entries
are: the number of bodies $N$, the initial value of $W_0$, the
primordial binary ratio $f$, time of core collapse $t_{cc}$, time of
dissolution $t_{dis}$ (computed at the point where only $2\%$ of the
initial mass of the system is left), core radius at core collapse
$r_c$, and core to half mass radius after core collapse
$r_c/r_h(t_{cc})$ (averaged for $5~t_{rh}(0)$ after core collapse),
fraction of binaries to singles in the core after core collapse
$(N_b/N_s)_c(t_{cc})$ (same average as for core to half mass radius), and
relative mass of the system at core collapse $M(t_{cc})/M(0)$.}
\label{tab:runs}
%%%%%%%%%%%%%%%%%%%%%%%%%%%%%%%%%%%
\begin{tabular}{llllllllll}
$N$ & $W_0$ & $f$ & $t_{cc}$ & $t_{dis} (98\%)$ & $\tilde{r}_c(t_{cc})$ & $r_c(t_{cc})$ &
 $r_c/r_h(t_{cc})$ & $(N_b/N_s)_c (t_{cc})$ & $M(t_{cc})/M(0)$  \\
\hline
 
%%%%%%%%%%%%%%%%%%%%%%%%%%%%%%%%%%%%%%%%%%%%%%%%%%%%%%%%%%%%%%%%%%%%%%%%%%%%
 512 &    7.0 &  10 &    1.6 &   81.2 &    0.147 &    0.161 &    0.179 & 0.35 &    0.99 (averaged)\\
1024 &    7.0 &  10 &    4.0 &   66.8 &    0.129 &    0.147 &    0.157 & 0.44 &    0.97 (averaged)\\ 
2048 &    7.0 &  10 &    5.6 &   59.7 &    0.108 &    0.128 &    0.143 & 0.47 &    0.92 (averaged)\\

\hline

 4096 &    3.0 &   0 &   11.9 &   17.3 &    0.019 &    0.023 &    0.081 &     0.04 &    0.26 \\
 4096 &    3.0 &   0 &   11.7 &   16.8 &    0.018 &    0.022 &    0.080 &     0.03 &    0.27 \\
 4096 &    3.0 &  10 &   11.1 &   13.8 &    0.071 &    0.075 &    0.210 &     2.84 &    0.15 \\
 4096 &    3.0 &  10 &   10.5 &   13.0 &    0.073 &    0.080 &    0.175 &     2.53 &    0.17 \\
 4096 &    3.0 &  20 &    9.5 &   13.4 &    0.085 &    0.097 &    0.230 &     4.48 &    0.23 \\
 4096 &    3.0 &  20 &   10.0 &   13.0 &    0.066 &    0.074 &    0.235 &     6.39 &    0.17 \\
 4096 &    3.0 &  50 &    9.8 &   13.9 &    0.075 &    0.090 &    0.190 &     8.84 &    0.24 \\
 4096 &    3.0 & 100 &   14.1 &   16.5 &    0.062 &    0.072 &    0.274 &    34.05 &    0.09 \\
\hline
 4096 &    7.0 &   0 &   11.5 &   50.3 &    0.015 &    0.022 &    0.030 &     0.02 &    0.79 \\
 4096 &    7.0 &   0 &   10.8 &   47.7 &    0.014 &    0.021 &    0.031 &     0.02 &    0.77 \\
 4096 &    7.0 &   2 &   20.8 &   51.8 &    0.035 &    0.045 &    0.041 &     0.03 &    0.52 \\
 4096 &    7.0 &   2 &   17.4 &   48.4 &    0.036 &    0.046 &    0.053 &     0.06 &    0.61 \\
 4096 &    7.0 &   5 &    4.1 &   52.2 &    0.091 &    0.110 &    0.129 &     0.24 &    0.95 \\
 4096 &    7.0 &   5 &   14.0 &   53.4 &    0.063 &    0.078 &    0.089 &     0.23 &    0.69 \\
 4096 &    7.0 &  10 &    3.3 &   48.0 &    0.100 &    0.121 &    0.132 &     0.61 &    0.95 \\
 4096 &    7.0 &  10 &    7.6 &   45.8 &    0.096 &    0.117 &    0.126 &     0.57 &    0.84 \\
 4096 &    7.0 &  20 &    2.0 &   41.4 &    0.132 &    0.158 &    0.132 &     1.24 &    0.98 \\
 4096 &    7.0 &  20 &    6.6 &   40.2 &    0.097 &    0.118 &    0.130 &     1.44 &    0.83 \\
 4096 &    7.0 &  50 &    7.5 &   44.0 &    0.113 &    0.141 &    0.163 &     4.24 &    0.84 \\
 4096 &    7.0 & 100 &    3.4 &   55.1 &    0.147 &    0.189 &    0.203 &    11.08 &    0.93 \\
\hline
 4096 &   11.0 &   0 &    0.0 &   28.7 &    0.019 &    0.028 &    0.034 &     0.02 &    1.00 \\
 4096 &   11.0 &   0 &    0.0 &   32.2 &    0.021 &    0.030 &    0.031 &     0.02 &    1.00 \\
 4096 &   11.0 &  10 &    0.0 &   25.1 &    0.082 &    0.103 &    0.121 &     0.54 &    1.00 \\
 4096 &   11.0 &  10 &    0.0 &   30.9 &    0.079 &    0.097 &    0.122 &     0.56 &    1.00 \\
 4096 &   11.0 &  20 &    0.0 &   23.8 &    0.074 &    0.094 &    0.128 &     1.08 &    1.00 \\
 4096 &   11.0 &  20 &    0.0 &   26.7 &    0.082 &    0.102 &    0.137 &     1.03 &    1.00 \\
 4096 &   11.0 &  50 &    0.0 &   28.4 &    0.080 &    0.101 &    0.158 &     2.86 &    1.00 \\
 4096 &   11.0 & 100 &    0.0 &   32.4 &    0.082 &    0.107 &    0.179 &    11.93 &    1.00 \\
\hline
 8192 &    7.0 &  10 &    7.1 &   38.7 &    0.074 &    0.092 &    0.108 &     0.77 &    0.82 \\
\hline
16384 &    3.0 &  10 &    9.4 &    9.6 &    0.034 &    0.038 &    0.080 &     6.05 &    0.02 \\ %    0.00
16384 &    3.0 &  20 &    8.5 &    9.2 &    0.036 &    0.043 &    0.241 &     6.75 &    0.05 \\ %    0.00
16386 &    7.0 &  10 &    7.2 &   32.4 &    0.069 &    0.086 &    0.097 &     1.16 &    0.76 \\ %    0.00
16384 &    7.0 &  20 &    6.2 &   28.5 &    0.076 &    0.096 &    0.095 &     2.53 &    0.81 \\ %    0.00
16384 &   11.0 &  10 &    0.0 &   20.1 &    0.056 &    0.074 &    0.102 &     0.78 &    1.00 \\ %    0.00
16384 &   11.0 &  20 &    0.0 &   19.3 &    0.069 &    0.090 &    0.105 &     1.77 &    1.00 \\ %    0.00

\hline
\hline

 512 &    7.0 &  10 &    1.8 &   83.9 &    0.137 &    0.153 &    0.196 & 0.35 &    0.99 (averaged + tidal cut-off)\\

 1024 &    7.0 &  10 &    3.5 &   75.4 &    0.124 &    0.142 &    0.172 &     0.39 &    0.96 (averaged + tidal cut-off)\\

 2048 &    7.0 &  10 &    4.7 &   67.2 &    0.111 &    0.132 &    0.146 &     0.53 &    0.94 (averaged + tidal cut-off)\\

 4096 &    7.0 &  10 &    3.5 &   56.3 &    0.095 &    0.118 & 0.129 &     0.64 &    0.94 (tidal cut-off)\\

 8192 &    7.0 &  10 &   10.9 &   50.4 &    0.077 &    0.096 &    0.108 &     0.74 &    0.78 (tidal cut-off)\\

16384 &    7.0 &  10 &    7.6 &   41.2 &    0.065 &    0.084 & 0.088 &  0.88 &    0.83 (tidal cut-off) \\

\hline

4096 &    3.0 &  10 &   12.9 &   17.0 &    0.061 &    0.068 &    0.275 &     2.30 &    0.22 (tidal cut-off)\\ %    0.01

8192 &    3.0 &  10 &   13.5 &   14.8 &    0.047 &    0.048 &    0.248
&  5.53  &    0.09 (tidal cut-off)\\ %    0.00 thalf=8.7

16384 &    3.0 &  10 &   14.2 &   15.3 &    0.041 &    0.046 &
0.272 &   8.31  &    0.07 (tidal cut-off)\\ %    0.00 thalf=9.0

\hline
\hline
%% Isolated W0=7 models, f=0.1 %%
  512 &    7.0 &  10 &    1.6 &   N/A &    0.141 &    0.157 &    0.193 &
0.32 &    1.00 (averaged, isolated) \\
 1024 &    7.0 &  10 &    3.2 &   N/A &    0.131 &    0.152 &    0.155 &
0.40 &    0.99 (averaged, isolated) \\
 2048 &    7.0 &  10 &    4.9 &   N/A &    0.112 &    0.134 &    0.128 &
0.50 &    0.99 (averaged, isolated) \\
 4096 &    7.0 &  10 &    5.6 &   N/A &    0.104 &    0.128 &    0.107 &
0.59 &    0.98 (averaged [5 sim.], isolated) \\
 8192 &    7.0 &  10 &    9.7 &   N/A &    0.082 &    0.104 &    0.094 &
0.64 &    0.97 (isolated) \\

\hline
\end{tabular}
\\ {\sl}
\end{table*}
\end{center}
%%%%%%%%%%%%%%%%%

\end{document}